\documentclass[useAMS,usenatbib]{mn2e}

\usepackage[english]{babel}
\usepackage[utf8]{inputenc}
\usepackage{amsmath}
\usepackage{graphicx}
\usepackage{graphics}
\usepackage{float}
\usepackage{array}
\usepackage{natbib}
\usepackage[utf8]{inputenc} 
\usepackage{mdwmath}
\usepackage{mdwtab}
\usepackage{multirow}

\newcommand{\be}{\begin{equation}}
\newcommand{\ee}{\end{equation}}
\newcommand{\ba}{\begin{eqnarray}}
\newcommand{\ea}{\end{eqnarray}}

\renewcommand{\vec}[1]{\boldsymbol{#1}}
\newcommand{\tens}[1]{\boldsymbol{\mathsf{#1}}}

\def\simless{\mathbin{\lower 2.5pt\hbox
{$\rlap{\raise 4.5pt\hbox{$\char'074$}}\mathchar"7218$}}}
\def\simgrt{\mathbin{\lower 2.5pt\hbox
{$\rlap{\raise 4.5pt\hbox{$\char'076$}}\mathchar"7218$}}}
\include{epsf}
\title[Adaptive Semi-linear Inversion]
{Adaptive Semi-linear Inversion of Strong Gravitational Lens Imaging}
\author[Nightingale \& Dye]
{\parbox{\textwidth}{J. W. Nightingale,$^{1}$\thanks{E-mail:ppxjwn@nottingham.ac.uk}
S. Dye$^{1}$
}
\vspace{4mm}\\
$^{1}$School of Physics and Astronomy, Nottingham University,
University Park, Nottingham, NG7 2RD, UK\\
}

\begin{document}

\bibliographystyle{mn2e}

\bibpunct{(}{)}{;}{a}{}{;}

\date{}

\pagerange{\pageref{firstpage}--\pageref{lastpage}} 

\pubyear{2015}

\maketitle

\label{firstpage}

\begin{abstract}

We present a new pixelized method for the inversion of gravitationally lensed extended source images which we term adaptive semi-linear inversion (SLI). At the heart of the method is an h-means clustering algorithm which is used to derive a source plane pixelization that adapts to the lens model magnification. The distinguishing feature of adaptive SLI is that every pixelization is derived from a random initialization, ensuring that data discretization is performed in a completely different and unique way for every lens model parameter set. We compare standard SLI on a fixed source pixel grid with the new method and demonstrate the shortcomings of the former when modeling singular power law ellipsoid (SPLE) lens profiles. In particular, we demonstrate the superior reliability and efficiency of adaptive SLI which, by design, fixes the number of degrees of freedom (NDOF) of the optimization and thereby removes biases present with other methods that allow the NDOF to vary. In addition, we highlight the importance of data discretization in pixel-based inversion methods, showing that adaptive SLI averages over significant systematics that are present when a fixed source pixel grid is used. In the case of the SPLE lens profile, we show how the method successfully samples its highly degenerate posterior probability distribution function with a single non-linear search. The robustness of adaptive SLI provides a firm foundation for the development of a strong lens modeling pipeline, which will become necessary in the short-term future to cope with the increasing rate of discovery of new strong lens systems.

\end{abstract}


\section{INTRODUCTION}\label{Intro}

Strong gravitational lensing has seen rapid progress over the past decade thanks to the advent of targeted searches for strongly lensed systems. Surveys such as the Sloan Lens ACS Survey (SLACS) \citep{Bolton2006, Auger2009}, Strong Lensing in the Legacy Survey (SL2S) \citep{Sonnenfeld2013b}, Sloan WFC Edge-on Late-type Lens Survey (SWELLS) \citep{Treu2011} and Baryon Oscillation Spectroscopic Survey (BELLS) \citep{Brownstein2012} have together found over a hundred strong galaxy-galaxy lenses. Of these observed systems, the source and lens galaxies span a range of redshifts, morphologies and environments and are thus beginning to bring unique insight to our understanding of galaxy structure and its evolution. With the number of observations set to significantly increase over the next decade, strong lensing will play an ever growing role in the foreseeable future of extra-galactic astronomy.

Accompanying this fast growing data set of strong lenses has been the development of a number of different methods for their modeling. These fall broadly into two categories depending on whether the source is modeled by a smooth parametric light profile or a discretized surface brightness distribution. Methods that fall within the former category tend to search over a fully non-linear parameter space spanned by both the lens and source parameters. Such methods have seen regular use in the literature, for example, in the analysis of both SLACS \citep{Bolton2008} and SL2S \citep{Gavazzi2012} where the method has been used to confirm the lensing nature of many systems and determine their Einstein radii. The fast run time and ease of use creates a niche for these fully non-linear methods but they lack sufficient accuracy to perform more complex lens modeling and they break down with irregular source morphology. Furthermore, owing to the typically large and complicated non-linear parameter space, these methods can not guarantee that the global best fit has been reached.

Methods using discretized source surface brightness distributions, which we will refer to hereafter as `pixelized methods', circumvent these shortcomings; reconstruction of the source using a pixel grid inherently accounts for the possibility of an irregular source morphology and makes calculation of the source light a linear problem \citep[][WD03 hereafter]{Warren2003}. This ultimately leads to an improved accuracy in fitting to the observed lensed image which subsequently enables lens modeling of greater complexity. However, pixelized methods can be computationally more expensive to run, especially when a high resolution source grid is used, and typically involve a greater investment of time to set up. These time demands have resulted in a mixture of fully non-linear and pixelized methods finding use in the literature, rather than sole application of the more sophisticated pixelized methods.

A striking omission among strong lensing studies is a demonstration of the reliability of pixelized methods with lens models that are more complex than simple isothermal density profiles. The singular power-law ellipsoid (SPLE) is one such model which gives rise to a more complex parameter space. The SPLE, with a volume mass density of the form $\rho$ $\propto$ $r^{-\alpha}$, where the power-law slope, $\alpha$, is a free parameter, is shown to be an excellent representation of the overall density profile of early type galaxies (ETGs) \citep{Koopmans2006, Barnebe2009}. Accordingly, the SPLE gives a significant improvement to the fitting of strong lensing data compared to a singular isothermal ellipsoid (SIE) profile. 

SLACS, BELLS and SL2S have made great progress in measuring $\alpha$ for over 100 strong lenses \citep{Koopmans2006, Bolton2012, Sonnenfeld2013b}. This work indicates that the observed total density profile of massive ETGs steepens with decreasing redshift, a measurement now being used to constrain galaxy formation models \citep{Oguri2014, Dutton2014}. However, determination of $\alpha$ was not made by fitting a SPLE to the strong lensing data, but instead by combining the velocity dispersion of the lens with the Einstein mass calculated from a SIE lens profile. This gives two measurements of a galaxy's mass at two different radii, which is combined through either an empirical mass scaling \citep{Bolton2008b} or kinematic analysis \citep{Treu2002, Sonnenfeld2013c}. Whilst this approach has its advantages, for example that an average slope is measured over a relatively wide range of radii, there are also limitations such as assumptions regarding lens mass sphericity when solving the Jeans equations and the fact that the error on the velocity dispersion usually dominates the uncertainty in $\alpha$ \citep{Koopmans2009}. 

In this paper, we advocate the use of purely strong lens data, on the basis that the stronger constraints arising from a full exploitation of the information contained in lensed images offsets the lack of dynamical data (and dispenses with the need for noisy kinematics). We present a new implementation of the semi-linear method of WD03, which we refer to as `adaptive semi-linear inversion' (adaptive SLI). We demonstrate several important improvements brought about by adaptive SLI over existing implementations. In particular, we concentrate on the application of adaptive SLI to the reconstruction of SPLE lens models and, crucially, we show how standard SLI introduces biases when measuring $\alpha$. We also advocate the use of the adaptive SLI method in a strong lensing reconstruction pipeline for application to large datasets from both existing surveys, e.g., SLACS, SL2S and SWELLS, and future surveys such as the Dark Energy Survey (DES) and Large Synoptic Survey Telescope (LSST) \citep{Oguri2010}.

Several other pixelized methods have been developed over the past decade which improve on the use of a regular Cartesian square grid. \citet{Dye2005} split up square pixels in high magnification regions to obtain a square grid adapted to the magnification, however the method retains the biases we describe. \citet{Tagore2014} make a number of improvements to this method in their {\tt pixsrc} program. \citet{Dye2014} perform lens modeling on multi-wavelength observations simultaneously, a feature we have implemented in adaptive SLI but not used in the present work. Adaptive SLI has the most in common with the adaptive grid of \citet{Vegetti2009}. We will show, however, how our different approach to source plane pixelization removes biases which are still present with their adaptive scheme.

The paper is laid out as follows: Section \ref{Method} describes the adaptive SLI method. We first describe how source plane pixelization is performed, followed by the linear regularization scheme and lens mass optimization. Section \ref{SqVsAd} presents a thorough comparison of the square and adaptive SLI methods with a focus on the biases inherent to the SPLE lens model with a square grid. We demonstrate how adaptive SLI removes these biases. Finally, a summary is given in section \ref{Discussion}.

\section{METHODOLOGY}\label{Method}

In this section we describe the adaptive SLI method. Section \ref{MethodNoReg} outlines source plane pixelization and inversion without regularization. We introduce regularization with a fixed lens model in Section \ref{MethodReg}. Section \ref{MethodMass} then deals with our approach to optimization of the non-linear lens parameters before concluding with a discussion of practicalities in Section \ref{MethodPrac}.

\subsection{Adaptive Source Plane Pixelization and Inversion}\label{MethodNoReg}

For a fixed set of lens model parameters, the SLI method of WD03 solves the linear problem of determining the surface brightnesses of source plane pixels such that coaddition of their individual lensed images provides the best fit to the observed lensed image. The goodness of fit is quantified by a merit function, $G$, which in the non-regularized case is simply $\chi^2$. The solution vector, $\vec{s}$, of source pixel surface brightnesses, $s_i$, is given by
\begin{equation}
\label{eqn:SNoReg}
\vec{s} = \tens{F}^{-1} \vec{D}
\end{equation}
where the square matrix $\tens{F}$ has elements $F_{ik} = \sum_{j=1}^{J}f_{ij}f_{kj}/\sigma_j^2$, the column vector $\vec{D}$ has elements $D_{i} = \sum_{j=1}^{J}f_{ij}d_{j}/\sigma_j^2$ and $f_{ij}$ is the $j$th pixel of the lensed PSF convolved image of source pixel $i$. $d_j \pm \sigma_j$ is the flux and statistical uncertainty of observed image pixel $j$. This gives a total of $I$ source pixels and $J$ lensed image pixels. In this unregularized case, for a square grid there may exist pixels unconstrained by the lens mapping, in which case $\tens{F}^{-1}$ may not exist.  We refer to the method of WD03 which uses a regular source pixel grid as `square SLI' hereafter.

As noted in WD03, discretization of the source plane is unrestricted. Adaptive SLI exploits this freedom in the way it allocates traced image pixels to source plane pixels. The spatial source plane co-ordinates of all $J$ traced image pixels are fed into an h-means clustering algorithm \citep{Hartigan1979}, which determines a set of `h-clusters'. An h-cluster is a region in the source plane to which a subset of image pixels is allocated. In this way, an h-cluster plays the role of a source plane pixel. Each h-cluster is then defined by its center coordinates which are found by minimizing the statistic $E$ given by
\begin{equation}
E = \sum^{I}_{i=1} e_i = \sum^{I}_{i=1} \sum^{K}_{k=1} {r_k}^2 \, .
\label{eqn:clusterE}
\end{equation}
$E$ is the sum of cluster `energies', where a cluster energy $e_i$ is the quadrature sum of the distances $r_k$ of each of its associated traced image plane pixels to the h-cluster center.

The h-means algorithm first calculates an initial set of h-cluster centers dependent on the source plane co-ordinates of all traced image pixels. This center initialization is randomized, such that a completely different set of centers will be calculated for a nearly identical set of co-ordinates. As we discuss later, this randomization is crucial in ensuring adaptive SLI addresses the discretization biases resulting from a fixed grid. The algorithm then proceeds by alternating between two processes; (i) for the current set of h-cluster centers, it allocates all traced image pixels to their nearest h-cluster center. (ii) for this new set of h-cluster assignments, all h-cluster centers are recalculated. This continues until either no point is moved or 20 iterations are performed. The resulting pixelization of the source plane is then dependent on the spatial distribution of traced image pixels and therefore also the magnification of the lens model, unlike the Cartesian grid used in square SLI and in a considerably more flexible way than the adaptive scheme of \citet{Dye2005}.

Each source plane h-cluster is the equivalent of a source plane pixel in the sense that all pixels in the image which belong to it are assigned the same surface brightness. However, unlike previous adaptive schemes, the pixels are completely arbitrary in shape and are not forced to adhere to any prescribed geometric forms which may bias the lens reconstruction. Furthermore, minimisation of cluster energies given in equation (\ref{eqn:clusterE}) ensures that the clusters are contiguous and do not overlap in spatial extent with neighbouring clusters. To simplify plotting of reconstructed sources with this scheme, in this paper we approximate clusters as Voronoi cells whose centers are the cluster centers. As described in the next section, this Voronoi gridding is also used to perform source plane regularization. However {\em we stress that while used for both visualization and regularization, the Voronoi grid itself is not a feature of the adaptive gridding scheme but rather only used after the adaptive grid is derived.}

We also employ sub-gridding of the image plane, which splits each image pixel into a set of square sub-pixels. The centers of these sub-pixels are then traced to the source plane for the clustering algorithm and inversion, rather than the centers of the full pixels. This increases the workload\footnote{The increase in clustering workload results in an insignificant increase to the duration of one full iterative step when optimizing lens model parameters.} of the clustering algorithm but removes pixel aliasing effects from the overall inversion which as we discuss later, are problematic when recovering lens model parameters. Note that this scheme achieves direct sub-gridding of the image unlike the reverse approach of \citet[][see their Appendix B]{Treu2004} who bilinearly interpolate the source plane for each full image pixel traced there. Throughout this paper, we divide image pixels into $4 \times 4$ sub-pixels except where otherwise stated.

In addition to sub-gridding, we also mask the observed image to remove background noise. The source plane maps only to points within this mask and therefore the goodness of fit is computed only for pixels within this mask. Masks were constructed to include all pixels affected by PSF smearing. The removal of background noise reduces the number of co-ordinates fed to the clustering algorithm. This ensures both greater efficiency and that a larger fraction of source plane pixels are dedicated to relevant regions in the image. We note that such tight masking can, in principle, result in additional extraneous images being masked out and thus incorrect solutions being deemed acceptable, although in practice this can be easily checked by computing final unmasked lens model images.

\subsection{Source Plane Regularization}\label{MethodReg}

Regularization adds an additional linear term, $G_L$, weighted by a scalar, $\lambda$, referred to as the regularization weight, to the merit function such that $G = \chi ^{2} + \lambda G_{L}$. In essence, this acts like a prior by more heavily penalizing reconstructed sources which are less smooth. In this way, regularization suppresses over-fitting to the image noise. The solution vector $\vec{s}$ of source pixel brightnesses is then
\begin{equation}
\label{eqn:SReg}
\vec{s} = [\tens{F}+\lambda \tens{H}]^{-1} \vec{D} \,\, ,
\end{equation}
where $\tens{H}$ is the regularization matrix which relates to the second derivative of $G_L$ as detailed by WD03. Unlike the unregularized case, for which $\tens{F}^{-1}$ may not exist, in this case $[\tens{F}+\lambda \tens{H}]^{-1}$ is guaranteed to exist for any sensible regularization scheme.

The regularity of a square grid makes regularization very straightforward, but with an adaptive grid, this is less so. We opt to use a Voronoi-neighbour based regularization scheme of the form
\begin{equation}
G_{L} = \sum_{i=1}^{I}\sum_{n=1}^{N_{v}} [s_i - s_{i,v}]^2 \,\, ,
\label{eqn:NNReg}
\end{equation}
where we use the h-cluster centers to determine the $N_{v}$ Voronoi neighbours for each cluster. Specifically, for each source plane pixel we find all neighbouring pixels with which it shares a Voronoi vertex, with this Voronoi grid derived from the h-cluster centers. This scheme then computes the difference in surface brightness between neighbouring source pixels, analogous to gradient regularization for a square grid.

The primary motivation for using this scheme is that it ensures regularization between pixels is evenly spread. We initially tested a nearest-neighbour scheme which regularized each source pixel with its three nearest neighbours. While this scheme still gave generally accurate results, it did not spread the regularization across source pixels evenly. For example while every pixel had three nearest neighbours with which they were paired to be regularized with, some pixels were the neighbour of more or less than three pixels. A consequence of this uneven spreading was that at caustic edges, where the cluster centers within the caustic are closer together, pixels were predominantly paired only to those also inside the caustic. This uneven spread of source regularization resulted in inaccurate source reconstruction at caustic edges and systematically offset lens parameter estimation. Our Voronoi scheme corrects this, pairing all pixels evenly and ensuring accurate source reconstruction and parameter estimation at caustic edges.

Different approaches to source plane pixelization have led to a variety of regularization schemes appearing in the literature. For example, the Voronoi grid of \cite{Vegetti2009} uses a scheme analogous to curvature regularization whilst \cite{Tagore2014} present multiple schemes for their square grid adaptive mesh, including one which imposes a Sersic light profile on source reconstruction. It is likely that different regularization schemes will suit different types of source morphology. The degree to which this effect influences lens model parameters is beyond the scope of the current work. Nevertheless, we find that our scheme outlined here is perfectly adequate for the test cases investigated in this work.

\subsection{Model Optimization}\label{MethodMass}

The SLI method was placed within a Bayesian framework by \citet[][S06 hereafter]{Suyu2006}, allowing the regularization weight to be set automatically by the data. Adaptive SLI uses this Bayesian wrapper in the form derived by \cite{Dye2008} and given by
\begin{eqnarray}
\label{eqn:evidence}
-2 \,{\rm ln} \, \epsilon &=& \chi^2
+{\rm ln} \, \left[ {\rm det} (\tens{F}+\lambda\tens{H})\right]
-{\rm ln} \, \left[ {\rm det} (\lambda\tens{H})\right]
\nonumber \\
& &
+ \lambda\vec{s}^{T}\tens{H\,s} + \sum_{k=1}^K\sum_{j=1}^{J_k}
{\rm ln} \left[2\pi (\sigma_j^k)^2 \right] \, ,
\end{eqnarray}
where $\epsilon$ is the Bayesian evidence which is maximized. 

There are three levels of inference in our model optimization. At the first level the model is assumed true (i.e., mass model and $\lambda$ are fixed) and we solve for the source surface brightnesses that best fit the observed image, as described in section \ref{MethodNoReg}. The second level finds the hyper parameter, $\lambda$, which maximizes the evidence for a given set of lens model parameters. This normalizes the posterior probability distribution in both lens and hyper parameters to give the most probable solution. \cite{Dye2008} set a second hyper parameter at this level, the `splitting factor', which determines the magnification threshold beyond which source pixels are split into finer pixels. In principle, we could introduce an additional hyper parameter to our adaptive SLI to mimic this behaviour, for example, the source grid resolution. However, since this reduces computational efficiency, we opt not to implement it in the present work. The third and final level of inference then maximizes $\epsilon$ to calculate the most probable lens model. This is a search of the posterior distribution of the lens parameters. We adopt the approximation of S06 by assuming that the probability distribution for $\lambda$ is a delta function which permits direct comparison of evidence between models. 
In the present work, we use this three tier approach for model optimization in the case of both square and adaptive SLI. In some instances, when demonstrating the effects of fixing the regularization weight, we instead minimize the more basic merit function $G = \chi^2 + \lambda G_L$. This is simply motivated by the fact that the other terms in the evidence, as given in equation (\ref{eqn:evidence}), remain constant in this case.

\begin{figure}
\centering
\includegraphics[width=0.5\textwidth]{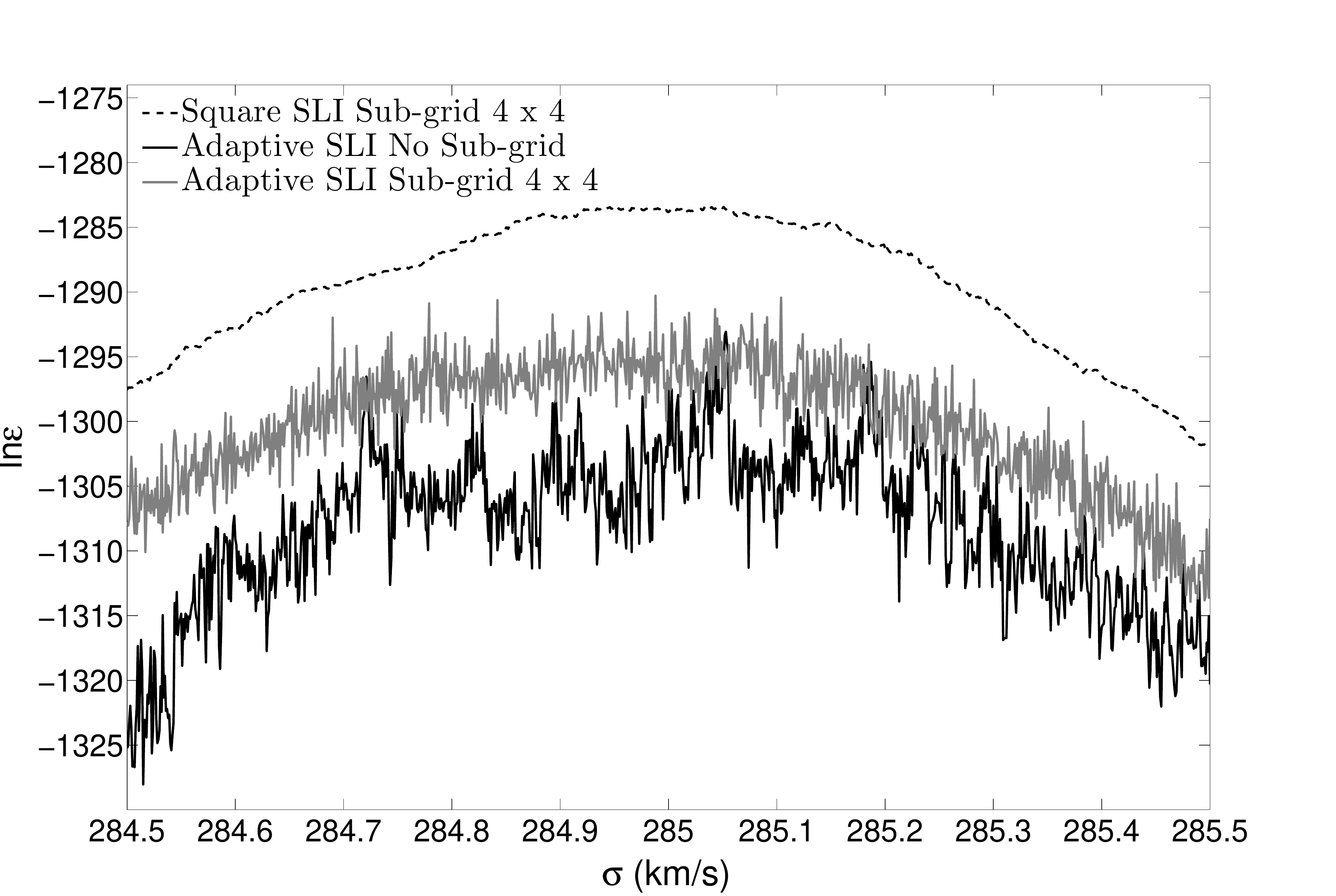}
\caption{Inversion of image 1 (see Figure \ref{figure:SynIm}) using the true lens model except the velocity dispersion, $\sigma$, which is varied over $\pm 0.5$\,kms$^{-1}$ about the true value. Square SLI (dashed line) gives a smooth and relatively noiseless evidence. Adaptive SLI (solid lines) gives rise to a much noisier and fluctuating evidence, owing to its constantly changing data discretization. The evidence is noisy both with (grey line) and without (black line) sub-gridding, although sub-gridding acts to slightly reduce the noise.}
\label{figure:LHNoise}
\end{figure}

\subsubsection{Data discretization}\label{DataDis}

Square SLI uses a fixed source plane grid, giving a non-linear parameter space which is smooth over small scales. A small perturbation to the lens parameters gives only a small change in $\epsilon$. This is because the source plane co-ordinates of traced image pixels remain nearly identical; the way the data is discretized, i.e., the allocations between image and source pixels also remains nearly identical. Source and image reconstruction then proceeds essentially unchanged, leading to only a fractional change in $\epsilon$.

With adaptive SLI, small perturbations to the lens parameters have a far more pronounced effect. Although the initial traced image pixel coordinates are nearly identical, their use in calculating the initial h-cluster centers is randomized. This means that even a tiny perturbation to these co-ordinates gives a different set of initial h-cluster centers. Given a different initialization, the clustering algorithm will then ultimately calculate a completely different set of final h-clusters and thus a completely unique source plane pixelization. Overall, source and image reconstruction remain similar given the lens model only changes minimally, however a relatively large change in $\epsilon$ is still possible. This is because for each lens model, adaptive SLI performs data discretization in a different and unique way, giving $f_{ij}$ matrices that are potentially very different.

Adaptive SLI is seeded such that identical lens parameters derive the same set of clusters, $\tens{F}$ matrix and therefore $\epsilon$. However, initialization is different for perturbations to the lens parameters of the order of just one part in $10^8$, a scale far smaller than that which model optimization probes. Therefore, in the context of a full non-linear search, {\em the inversion of every lens model uses a set of h-clusters which always discretize the data in a different, unique and unrelated way}. 

The notion of data discretization being unrelated to previous discretizations turns out to be vitally important. This property is not present in previous methods since in these methods, pixelization is computed in a manner that is deterministic and/or smoothly varying with lens model parameters. A consequence of this for our adaptive scheme, however, is that the non-linear parameter space is very noisy, making determination of convergence and parameter marginalization challenging. 

This is illustrated in Figure \ref{figure:LHNoise}. The figure shows the variation of evidence with velocity dispersion, $\sigma$, computed using the setup for simulated image 1 described later, keeping all other lens parameters fixed. We show this variation for square SLI and adaptive SLI with and without image sub-gridding. Square SLI gives rise to a relatively smooth evidence surface as $\sigma$ is varied. However, in the adaptive SLI case, large jumps in $\epsilon$ occur even over the very small steps in $\sigma$ of 1\,ms$^{-1}$ plotted, as a result of the unrelated data discretization with each step. Image sub-gridding slightly lessens the size of the jumps by largely removing pixel aliasing effects but the discretization effect remains present.

\subsection{Optimization Practicalities}\label{MethodPrac}

Our non-linear search must be suited to sampling this noisy and fluctuating parameter space. This is something traditional MCMC searches are not equipped to deal with, owing to their reliance on a walk up a relatively smooth likelihood surface. Therefore, we instead use the {\tt MultiNest} algorithm \citep{Feroz2007,Feroz2009}, which implements the nested sampling Monte Carlo technique of \citet{Skilling2006}. The algorithm initially generates a set of live points within parameter space which probe the smoother large scales to map out the high evidence regions. The lowest evidence points are subsequently replaced iteratively, resulting in convergence towards high evidence regions where noise slowly becomes more prominent. We find that this is a very efficient way to cope with the noise within our parameter space and with a sufficient number of live points, {\tt MultiNest} accurately determines the most probable solutions. We use {\tt MuiltNest} to perform model optimization with both square and adaptive SLI, the latter requiring comparatively more iterations to complete given its noisy parameter space.

A well recognized problem when performing model optimization of strong lens data is the existence of unwanted solutions which correspond to either an over or under estimated lens mass. In both cases, the reconstructed source resembles a demagnified version of the observed lensed image rather than a much more compact source at the correct solution. While regularization ensures that the evidence of these solutions is well below that of the global maximum, they occupy a large volume of parameter space which {\tt MultiNest} can waste significant time exploring. We therefore apply a set of coupled priors which ensure that these incorrect solutions are not accessible to {\tt MultiNest}. We determine these with a grid search over the lens parameters $\alpha$, the axis ratio, $q$, and the velocity dispersion, $\sigma$, to identify the distinct and isolated regions where the incorrect solutions lie. Initial {\tt MultiNest} sampling then uses randomized triplets of $\sigma$, $q$ and $\alpha$ drawn from this region, with care taken to ensure the entire volume of this region is sampled and no viable solutions are trimmed or lost. After {\tt MultiNest} has run for a while and achieved a specific accuracy, it switches to elliptical sampling mode where the current live points create an ellipsoidal sampling contour in parameter space and $\sigma$, $q$ and $\alpha$ are then drawn instead from these contours. 

Once sampling is complete, the set of all accepted points, including the current live points, then map out the evidence surface in parameter space. Each parameter is marginalized over in one dimension to calculate its posterior probability distribution, of which the median is computed to give final parameter estimates. Errors presented correspond to $1\sigma$ confidence bounds, i.e., the 16th and 84th percentile of the posterior probability distribution.

\section{COMPARISON OF SQUARE AND ADAPTIVE SLI}\label{SqVsAd}

\begin{figure*}
\centering
\includegraphics[width=1.0\textwidth]{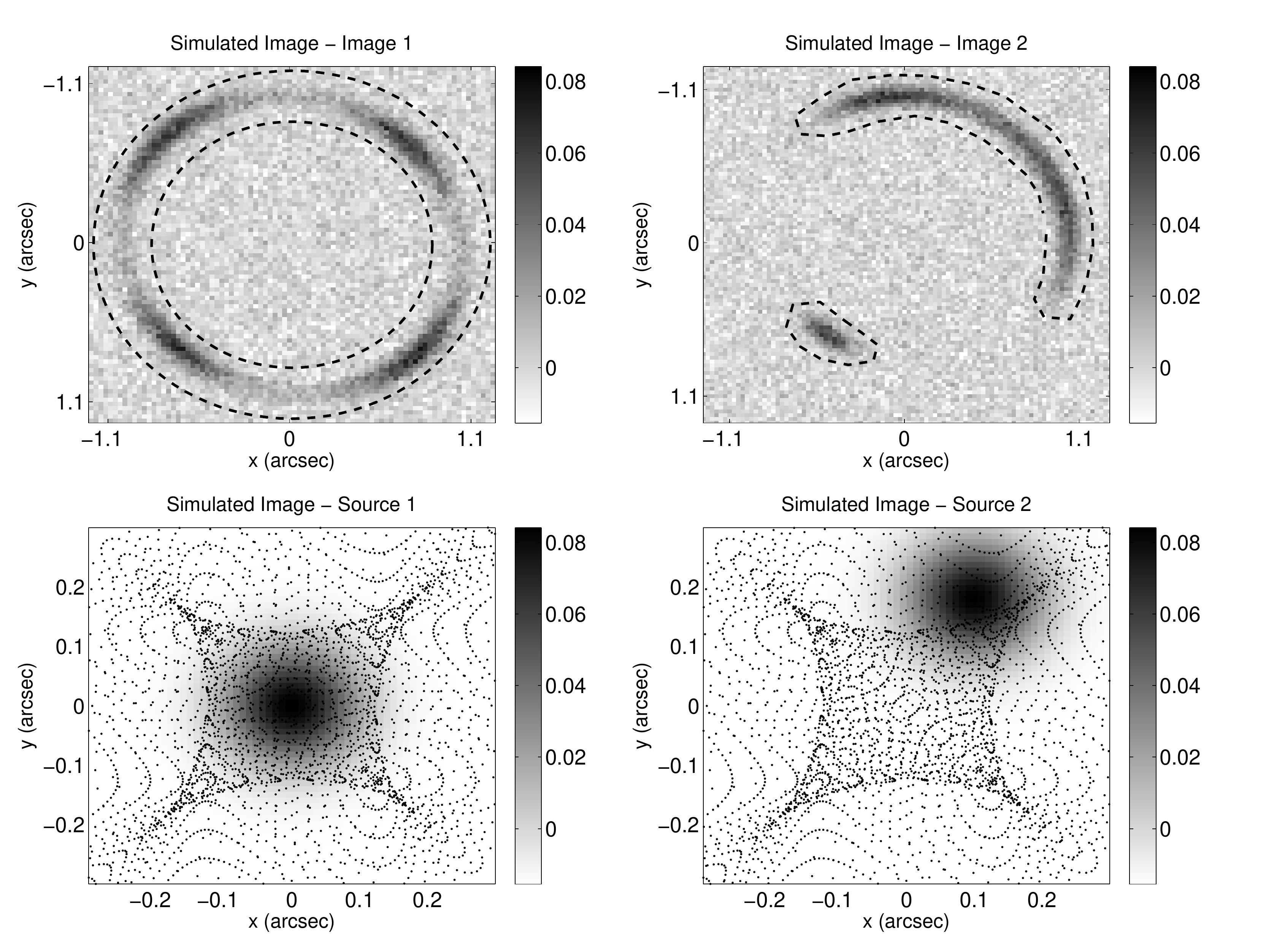}
\caption{{\em Top row} - Simulated images 1 and 2 generated with the same input SPLE lens model (lens x and y offsets = 0, $\sigma = 285$ km/s, q (b/a) = 0.8, $\phi = 45\,^{\circ}$ and $\alpha = 2.0$). Images have a pixel scale $0.048"$, are convolved with a Gaussian PSF with FWHM = $0.13"$ and have S/N=116 and 104. Only pixels within the image masks, plotted with a dashed line, are used during an inversion. {\em Bottom row} - Simulated sources of images 1 and 2 created using a symmetric Gaussian with FWHM = 0.071", located centrally for image 1 (x = y = 0") and on the top right caustic cusp for image 2 (x = 0.1", y = 0.18"). The caustic and magnification are illustrated by the black dots which are traced image pixels for the entire image plane using the input lens model.}
\label{figure:SynIm}
\end{figure*}

In this section, we demonstrate the advantages of the adaptive SLI method over conventional square SLI. In particular, we emphasise a variety of reconstruction biases inherent with square SLI which adaptive SLI eliminates. Our comparison makes use of the SPLE lens model which has a volume mass density profile of the form 
\begin{equation}
\rho(r) = \rho_0 (r/r_0)^{-\alpha},
\label{eqn:SPLE}
\end{equation}
with a variable power-law index, $\alpha$, and fixed normalization $\rho_0r_0^\alpha$. We also use a SIE lens profile achieved by fixing $\alpha=2$.

Deflection angles are computed using the formalism of \cite{Keeton2002} which uses an equivalent velocity dispersion parameter, $\sigma$, for lens mass normalization, relating to the Einstein radius, $b$, by $\sigma = \sqrt{(b c^2 / 4 \pi) (D_{os}/D_{ls})}$, where $D_{os}$ and $D_{ls}$ are the angular diameter distances from the observer to the source and from the lens to the source respectively. The SIE model has a total of five free parameters; the co-ordinates of the lens center, $(x,y)$, the velocity dispersion, $\sigma$, the axis ratio, $q$ (semi-minor axis/semi-major axis), and lens rotation angle, $\phi$, defined counter-clockwise from the $y$ axis. The SPLE model has the additional sixth parameter, the density slope, $\alpha$. 

We use two synthetic lensed images in our comparison. These are generated using a SPLE lens model with parameters $(x,y)=(0,0)$, $\sigma = 285$ km/s, $q = 0.8$, $\phi = 45\,^{\circ}$ and $\alpha = 2$. Both images have a pixel scale of $0.048"$ and are convolved with a Gaussian point spread function (PSF) with a full width at half maximum (FWHM) of $0.13"$. A Gaussian noise map to mimic fixed read noise is added to both images, giving a total signal to total noise ratio of 116 and 104 in the masked regions of image 1 and 2 respectively. Sources are modeled as 2D symmetric Gaussians with FWHM = 0.071". The source of image 1 is centered exactly on the optic axis, giving an Einstein cross image, whereas in image 2, the source is positioned just above the top right cusp of the inner caustic, giving a standard cusp-caustic image. We place the source at a redshift of $z = 3.0$, the lens at a redshift of $z = 0.3$ and the cosmological parameters assumed are $\Omega_{m} = 0.3$, $\Omega_{\Lambda} = 0.7$ and $h=0.7$. Figure \ref{figure:SynIm} shows both simulated images, their masks and corresponding simulated sources. 

It is likely that accurate modeling of these images is more challenging than the majority of real lensed images. As \cite{Lagattuta2012} discuss, the typically more irregular distribution of light in a real source gives rise to a less degenerate lensed image which in turn allows tighter constraints on the lens modeling. Moreover, our S/N and image resolution are selected to be lower than that presently achieved in the highest quality strong lensing data sets. In addition, multi-wavelength observations of strong lenses are becoming commonplace and through their simultaneous analysis, offer lens modeling of even greater accuracy \citep[e.g.,][]{Dye2014}. Our results therefore offer a fairly conservative view of what can be achieved with higher quality and more comprehensive imaging but nevertheless provide a suitable basis upon which to make our comparison of inversion methods.

\subsection{Source Plane Pixelization}\label{SrcPlanePix}
\begin{figure*}
\centering
\includegraphics[width=1.0\textwidth]{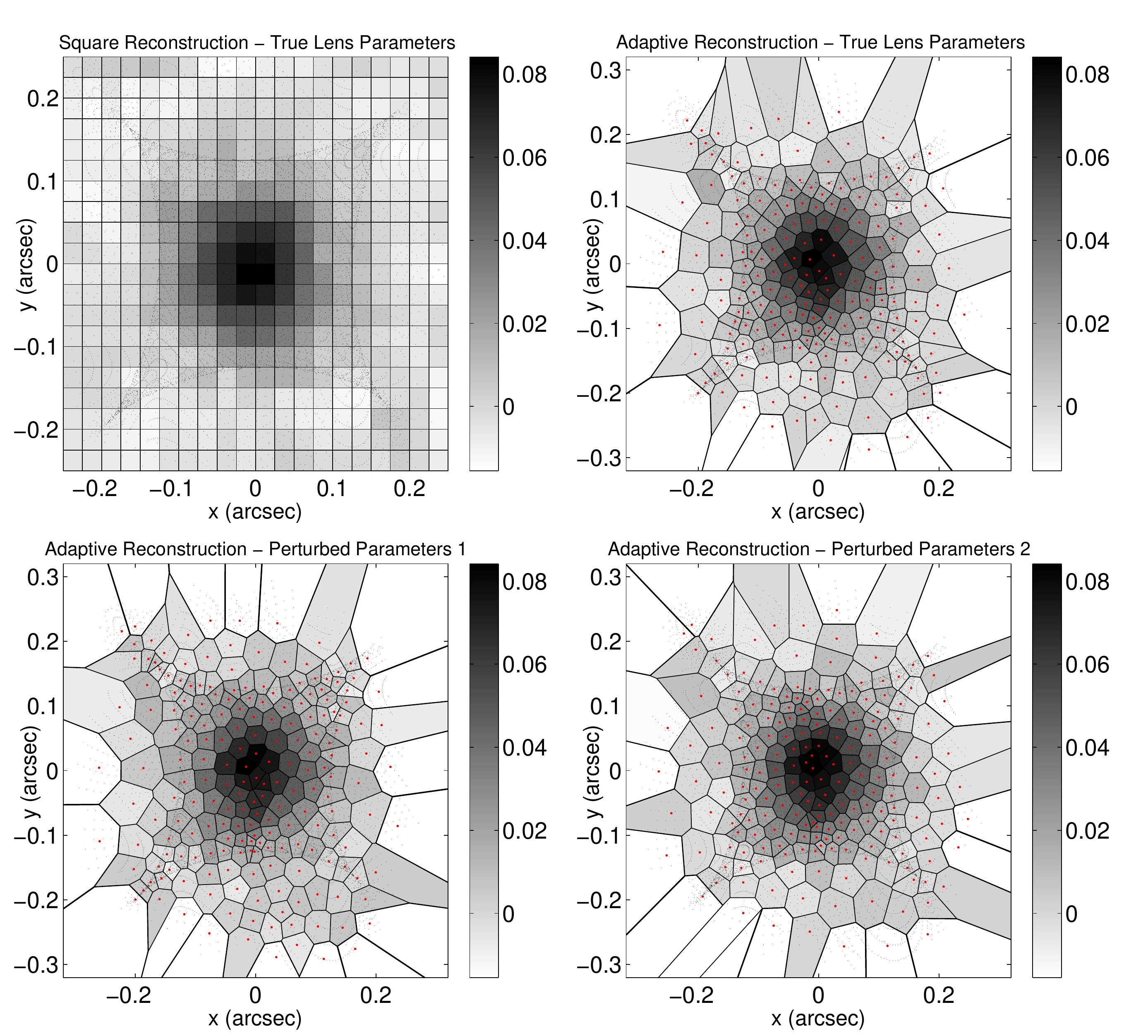}
\caption{{\em Top row} - A comparison of square and adaptive SLI pixelizations for image 1 using the input lens model. {\em Bottom row} - two additional adaptive SLI pixelizations using the input lens model but with $\sigma$ perturbed by +1 m/s (left) and -1 m/s (right). The reconstructed source for each inversion is overlaid. The square grid has a resolution of $20 \times 20$ pixels and size $0.5'' \times 0.5''$. The adaptive grid uses a resolution of 200 pixels and an arbitrarily large size to contain all traced image pixels (but shown here at $0.64'' \times 0.64''$ for comparison). All four inversions use image subgridding of $4 \times 4$; however to reduce the image size each black dot shows the traced location of image sub-pixels derived for subgridding of size $2 \times 2$. Red dots correspond to each h-cluster center. Note that source pixels are more organically shaped than the Voronoi cells which we display purely for clarity.}
\label{figure:SAComp}
\end{figure*}

The top row of figure \ref{figure:SAComp} shows the source plane pixelization and reconstruction of image 1 using the input lens model for both square and adaptive SLI. There are clear disadvantages with the use of a fixed grid in square SLI: (i) The grid is set up prior to the inversion, meaning that both the grid size and position are not determined by the lens model and that it may align poorly with the intrinsic distribution of source light; (ii) The fixed pixel area results in a highly varying magnification between source pixels such that a sub-set of pixels will dominate the inversion and unevenly spread the uncertainty; (iii) Outer source pixels may not map to any image pixels and are then constrained solely by regularization. 

By contrast, adaptive SLI's pixelization matches the lens model magnification and completely removes the need to specify a source plane size or location. Furthermore, uncertainty between source pixels is more evenly spread and there are no source pixels constrained solely by regularization. 

The bottom row of figure \ref{figure:SAComp} shows two additional source reconstructions performed on the adaptive grid. Both use the same lens model as before apart from a tiny perturbation to the velocity dispersion of $\pm1$\,m/s (i.e., a fractional change of $3.5\times10^{-6}$). All three adaptive pixelizations are globally similar, as expected given the almost identical magnification pattern, but upon closer inspection, it is apparent that the source pixels have significantly different centers, sizes and shapes. We stress again that we use Voronoi cells to simplify plotting whereas the underlying source pixelization is set by clusters of traced image pixels. The figure serves to illustrate the effect described in section \ref{DataDis}, whereby adaptive SLI derives a completely different set of h-clusters for every lens model and therefore always discretizes the source plane data in a unique and unrelated way. 

\subsection{SIE Lens Parameter Estimation}\label{SIE_Comp}

\begin{table*}
\begin{tabular}{ l | l | l | l | l | l | l }
\multicolumn{1}{p{0.8cm}|}{\centering Image} 
& \multicolumn{1}{p{0.8cm}|}{\centering Method} 
& \multicolumn{1}{p{0.7cm}|}{\centering $x$ \\ (arcsec)} 
& \multicolumn{1}{p{0.7cm}|}{\centering $y$ \\ (arcsec)}
& \multicolumn{1}{p{0.4cm}|}{\centering $\sigma$ \\ (km/s) } 
& \multicolumn{1}{p{1.2cm}|}{\centering $q$ \\ $(=b/a)$}
& \multicolumn{1}{p{0.9cm}}{\centering $\phi$ $(\,^{\circ})$} 
\\[-8pt] \hline
& & & & & & \\[-6pt]
1 & adaptive & $-0.0003 ^{+0.0015} _{-0.0020} $ & $ 0.0010 ^{+0.0016}_{-0.0020} $ & $ 284.930 ^{+0.260}_{-0.197} $ & $ 0.8003 ^{+0.0022}_{0.0026} $ & $ 44.578 ^{+0.166}_{-0.186} $ \\[2pt]
1 & square & $ 0.0001 ^{+0.0023} _{-0.0016} $ & $ 0.0003 ^{+0.0022}_{-0.0023} $ & $ 285.057 ^{+0.253}_{-0.284} $ & $ 0.7991 ^{+0.0027}_{0.0028} $ & $ 44.648 ^{+0.213}_{-0.228} $ \\ \hline
& & & & & & \\[-6pt]
2 & adaptive & $-0.0009 ^{+0.0038} _{-0.0031} $ & $-0.0043 ^{+0.0067}_{-0.0080} $ & $ 285.550 ^{+0.449}_{-0.449} $ & $ 0.7954 ^{+0.0042}_{0.0043} $ & $ 45.709 ^{+0.514}_{-0.564} $ \\[2pt]
2 & square & $ 0.0001 ^{+0.0041} _{-0.0045} $ & $-0.0063 ^{+0.0064}_{-0.0077} $ & $ 285.740 ^{+0.551}_{-0.574} $ & $ 0.7932 ^{+0.0054}_{0.0055} $ & $ 45.540 ^{+0.549}_{-0.524} $ \\
\end{tabular}
\caption{Estimated SIE lens parameters for both square and adaptive SLI using images 1 and 2. Square SLI has a source plane resolution of $20 \times 20$ and size $0.5'' \times 0.5''$. Adaptive SLI uses 200 source pixels. Both inversions use image subgridding of $4 \times 4$. Both methods accurately model a SIE lens profile.}
\label{table:SIE}
\end{table*}

Our next aim is to investigate how well the square and adaptive SLI methods can recover the parameters of the input lens model used to generate our simulated images. In this section, we consider the simpler case of the SIE model, i.e., we fix the density slope to $\alpha=2$ to match the input value.

In this case, the non-linear search has five free parameters which we limit with top hat priors to reduce the search volume of parameter space: $x$ and $y$ (each with priors $-0.05''$ to $0.05''$), $\sigma$ (prior 260 km/s to 305 km/s), $q$ (prior 0.7 to 0.9) and $\phi$ (prior 40$\,^{\circ}$ to 50$\,^{\circ}$). For the square SLI, we set the source plane grid to a size of $0.5'' \times 0.5''$ with $20 \times 20$ pixels. The adaptive SLI grid has an arbitrarily large source plane size and we use 200 adaptive pixels. Both inversions use image subgridding of $4 \times 4$. We check all results to ensure that none has a solution near a prior edge. 

The results are shown in Table \ref{table:SIE}. As the table shows, both methods estimate all parameters correctly with similar errors for both images. 

\begin{figure*}
\centering
\includegraphics[width=1.0\textwidth]{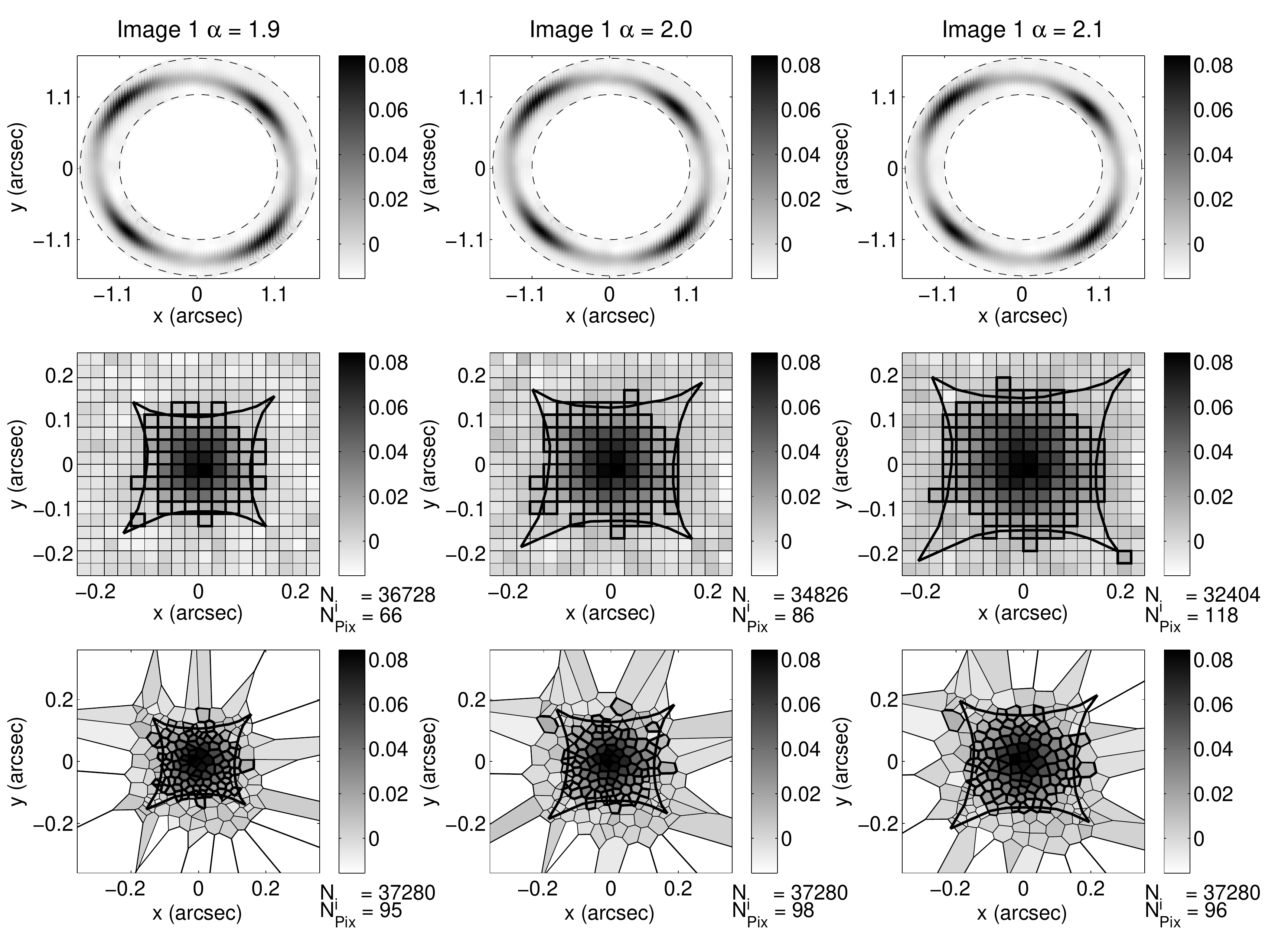}
\caption{{\em Top row} - reconstruction of image 1 with square SLI using the most likely lens models given in table \ref{table:SPLE} with $\alpha$ fixed to 1.9, 2.0 and 2.1. {\em Middle row} - the corresponding reconstructed source obtained from square SLI (resolution $20 \times 20$, size $0.5'' \times 0.5''$; caustic overlaid). {\em Bottom row} - The same lens reconstruction but using adaptive SLI (200 source pixels). Both inversions use image sub-gridding of $4 \times 4$. The geometric scaling of the source plane with $\alpha$ is immediately clear. $N_{\rm i}$ is the number of traced sub-gridded image pixels within the source plane. $N_{\rm Pix}$ is the number of source pixels with a count above $1.5\times$ the background noise, highlighted by a bold pixel edge.}
\label{figure:SPLE}
\end{figure*}

\subsection{SPLE Lens Parameter Estimation}\label{SPLE_Degen}

In this section, we apply square and adaptive SLI to our simulated images with the more general SPLE lens profile. We use the same inversion setup as in Section \ref{SIE_Comp} and the same priors on the lens position and $\phi$. We use a tophat prior on $\alpha$ over the range 1.75 to 2.25 and we calculate $\sigma$ and $q$ using randomized triplets as described in section \ref{MethodPrac}. 

Our initial run using square SLI finds $\alpha = 2.041 ^{+0.026} _{-0.037} $ for image 1 and $\alpha = 2.114 ^{+0.0585} _{-0.0530}$ for image 2. Clearly, both results are inconsistent with the input lens model. As we discuss below, this failure is due to degeneracies within the SPLE lens profile and biases within the square SLI method itself. 

In terms of degeneracies, the SPLE model has the well documented mass-slope degeneracy, whereby a more centrally concentrated mass distribution (i.e., higher $\alpha$) and a lower overall lens mass normalization produces a similar lensing effect (and vice versa). The net result of this is a geometric scaling of the source plane, such that the source expands for increasing $\alpha$. This is similar to both the fully degenerate mass-sheet transformation \citep{Schneider2013} and source plane transformation \citep{Sluse2013}, where a transformation of the lens mass alongside a geometric scaling of the source plane produces an identical set of observables. As shown in \cite{Schneider2013} and  \cite{Sluse2013}, both these degeneracies are formally broken for the specific case of a SPLE lens, ensuring that for our simulated images the correct lens model corresponds to a unique solution, which can be measured providing modeling is performed accurately. We stress that these degeneracies are broken only because we know a SPLE model was used to create our lensed images. However a strong degeneracy is still present and our lens parameterisation is such that the degeneracy is also dependent on the axis ratio parameter, $q$. We refer to this three-way degeneracy as the '$\sigma - q - \alpha$ degeneracy' hereafter.

\begin{table}
\begin{tabular}{ l | l | l }
\multicolumn{1}{p{0.4cm}|}{\centering $\sigma$ \\ (km/s)} 
& \multicolumn{1}{p{1.2cm}|}{\centering q (b/a)}
& \multicolumn{1}{p{0.9cm}}{\centering $\alpha$} 
\\ 
\hline
296.841 & 0.8381 & 1.9 \\ 
285.033 & 0.8010 & 2.0 \\
276.286 & 0.7564 & 2.1 \\
\end{tabular}
\caption{The most likely parameters determined by square SLI on image 1 for a SPLE lens model with $\alpha$ fixed to 1.9, 2.0 or 2.1. These are the parameters used to plot the image and source reconstructions in figure \ref{figure:SPLE}.} 
\label{table:SPLE}
\end{table}

To illustrate this degeneracy we fit a SPLE lens profile to image 1 using square SLI, with $\alpha$ fixed to 1.9, 2.0 and 2.1. The most likely solution for each is given in table \ref{table:SPLE} and the corresponding image and source reconstructions are shown in figure \ref{figure:SPLE}. The top row of figure \ref{figure:SPLE} shows the lensed reconstructed source of each solution, where it is immediately clear that solutions within this degeneracy produce near-identical images. In the middle and bottom rows of the figure, the geometric source scaling is clearly visible.

This poses a significant challenge to lens modeling. Although square SLI has the sensitivity to find the maximum evidence, the presence of even minor systematic biases within this degeneracy will push the solution a long way from the true parameter set. Such biases arise due to the arbitrary manner in which the square source plane is gridded as well as features inherent to the method itself. In the following subsections, we describe these biases in detail and demonstrate how adaptive SLI removes them. 

\subsubsection{Bias 1: The number of traced image pixels}\label{NiBias}

\begin{figure}
\centering
\includegraphics[width=0.5\textwidth,height=0.44\textwidth]{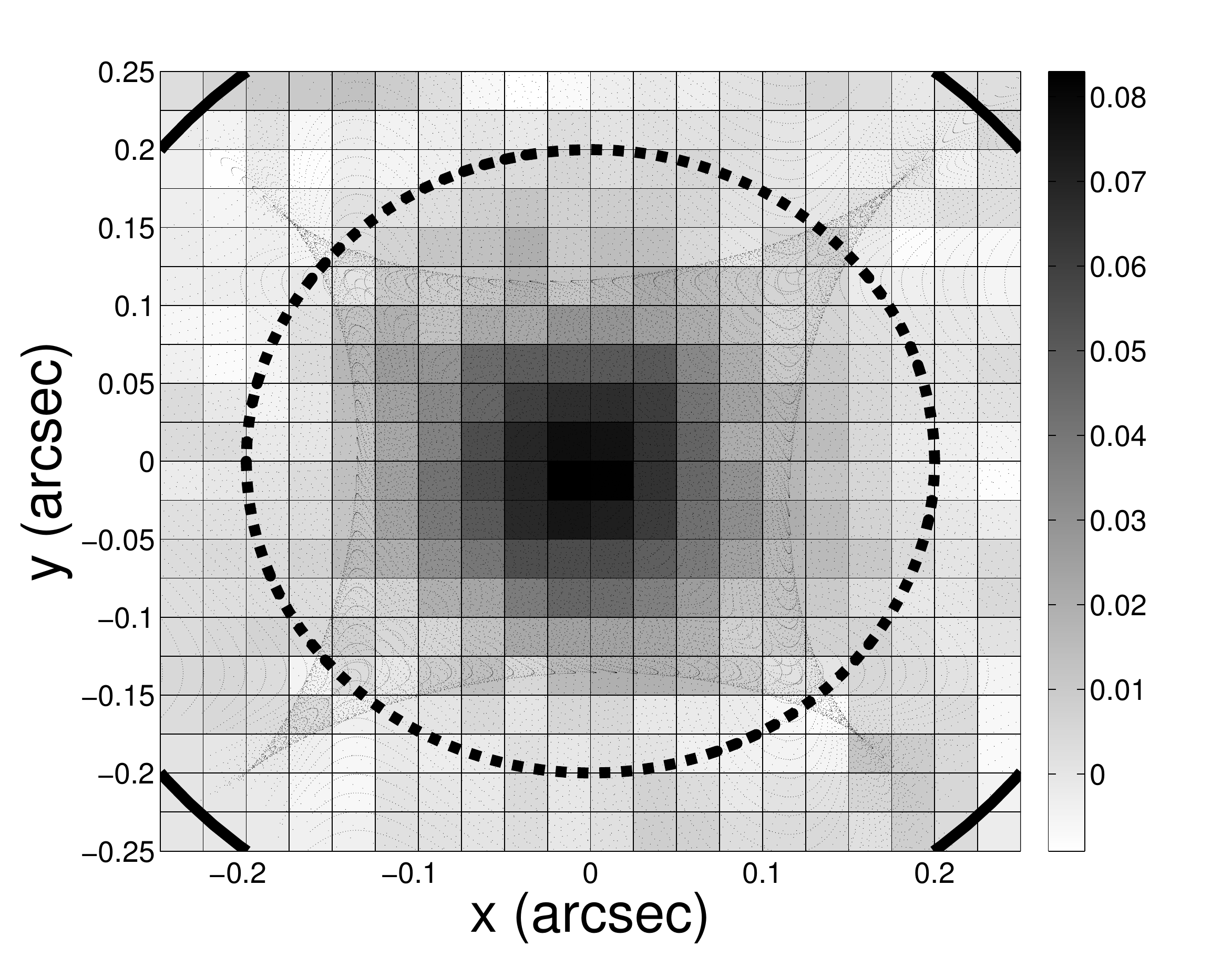}
\caption{Reconstruction of image 1 with the input lens model. To investigate the effect of a changing the number of traced image pixels, $N_i$, a circular mask is placed around source reconstruction and only points within the mask are then used by the inversion. The mask shown by the thick line corresponds to the smallest decrease in $N_i$ and that shown by the dashed line corresponds to the largest decrease in $N_i$. All other masks have intermediate radii thus gradually reducing the value of $N_i$. All points omitted correspond to background noise in the observed image.}
\label{figure:NiRings}
\end{figure}

In our preceding square SLI example, we used a source plane size of $0.5'' \times 0.5''$ as a compromise between being sufficiently large to encompass the source light and small enough not to compromise the source plane resolution. A consequence of this is that a significant fraction of image pixels trace back to locations outside the source plane. Some of these image pixels which trace to just outside the edge of the source plane still constrain the source due to PSF convolution and sub-gridding, but many will not, and yet all image pixels contribute to the $\chi^2$ term in equation (\ref{eqn:evidence}). As \citet{Vegetti2009} point out, as the lens model is iterated during optimization, the number of image pixels which trace to a point within the defined source plane area can vary. The resulting variation in the number of degrees of freedom (NDOF) causes problems for model inference if not taken into account. 

To explore the effect of varying the NDOF, we use a square souce grid with a fixed SPLE model and vary the number of image sub-pixels, $N_i$, that trace to within a circular source plane aperture centerd on the reconstructed source. By varying the radius of this aperture and ignoring traced image pixels outside it, (i.e., we set a null image for all exterior source pixels in $f_{ij}$ described in Section \ref{MethodNoReg}), we vary $N_i$. The number of source pixels are kept constant so as not to contribute to the changing NDOF; those pixels not within the aperture remain constrained by regularization. This mimics the effect of image pixels tracing outside a regular source plane but without the added complication of varying the parameterization of the problem.

\begin{figure}
\centering
\includegraphics[width=0.5\textwidth]{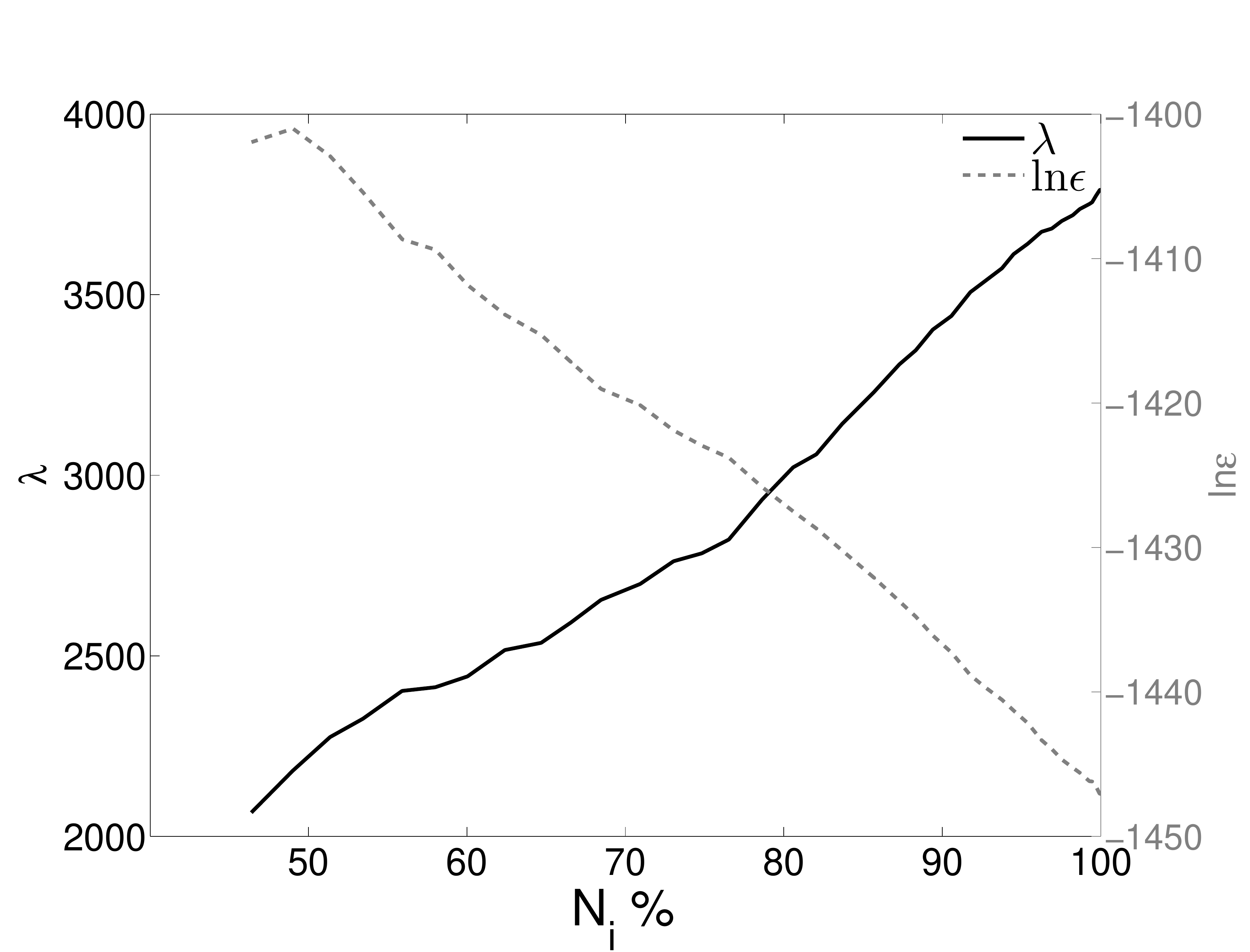}
\caption{Dependence of the regularization coefficient $\lambda$ (thick black) and evidence, $\epsilon$, (dashed grey) on the number of traced image pixels $N_i$. A reduction in $N_i$ results in a lower $\lambda$ being set and higher overall value of $\epsilon$.}
\label{figure:NiRegLH}
\end{figure}

\begin{figure}
\centering
\includegraphics[width=0.47\textwidth]{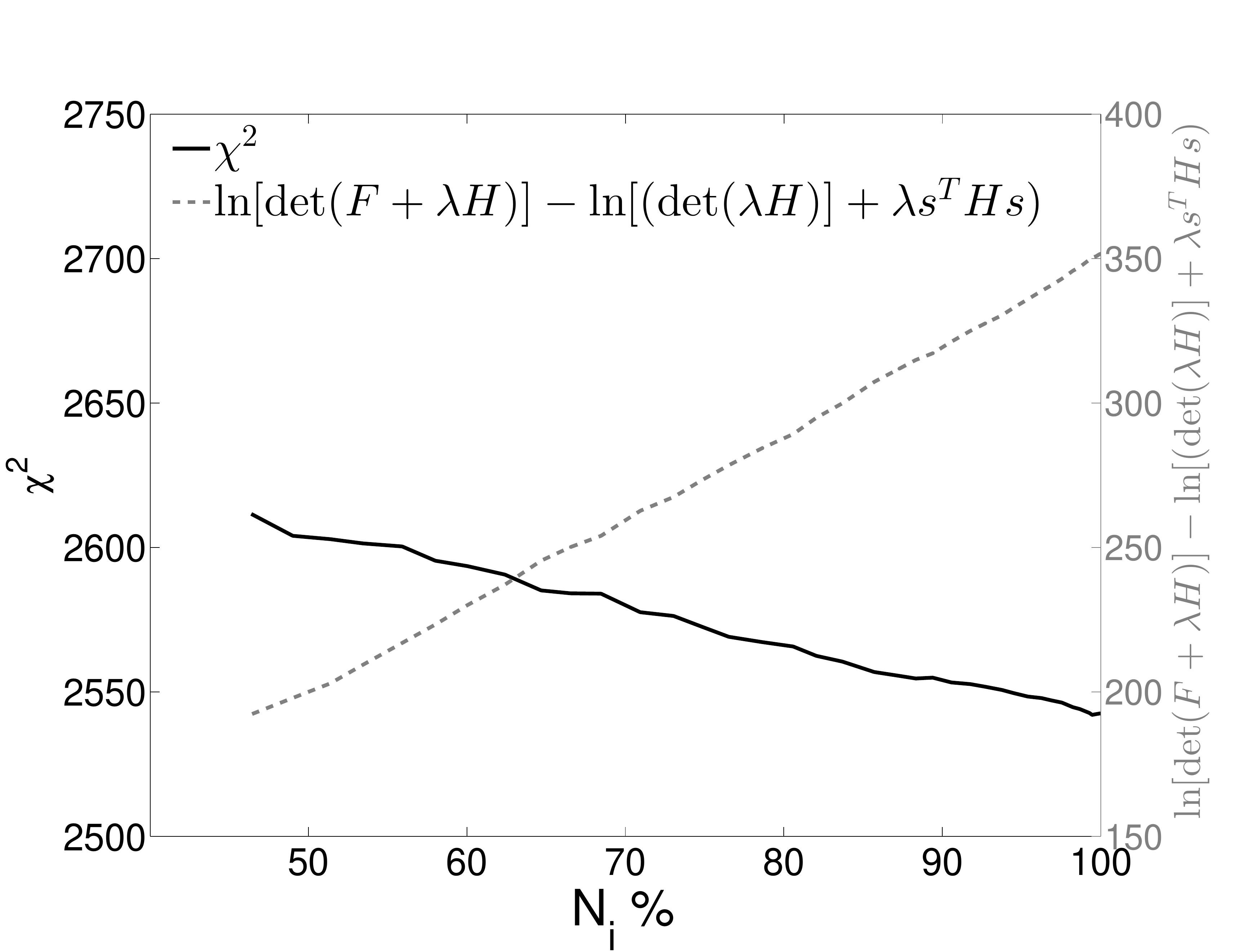}
\caption{Dependence of the image residual subtraction $\chi^2$ (thick black) and evidence regularization terms (dashed gray) on the number of traced image pixels $N_i$. A reduction in $N_i$ results in a worse residual fit and lower regularization terms.}
\label{figure:NiChiReg}
\end{figure}

\begin{figure}
\centering
\includegraphics[width=0.5\textwidth]{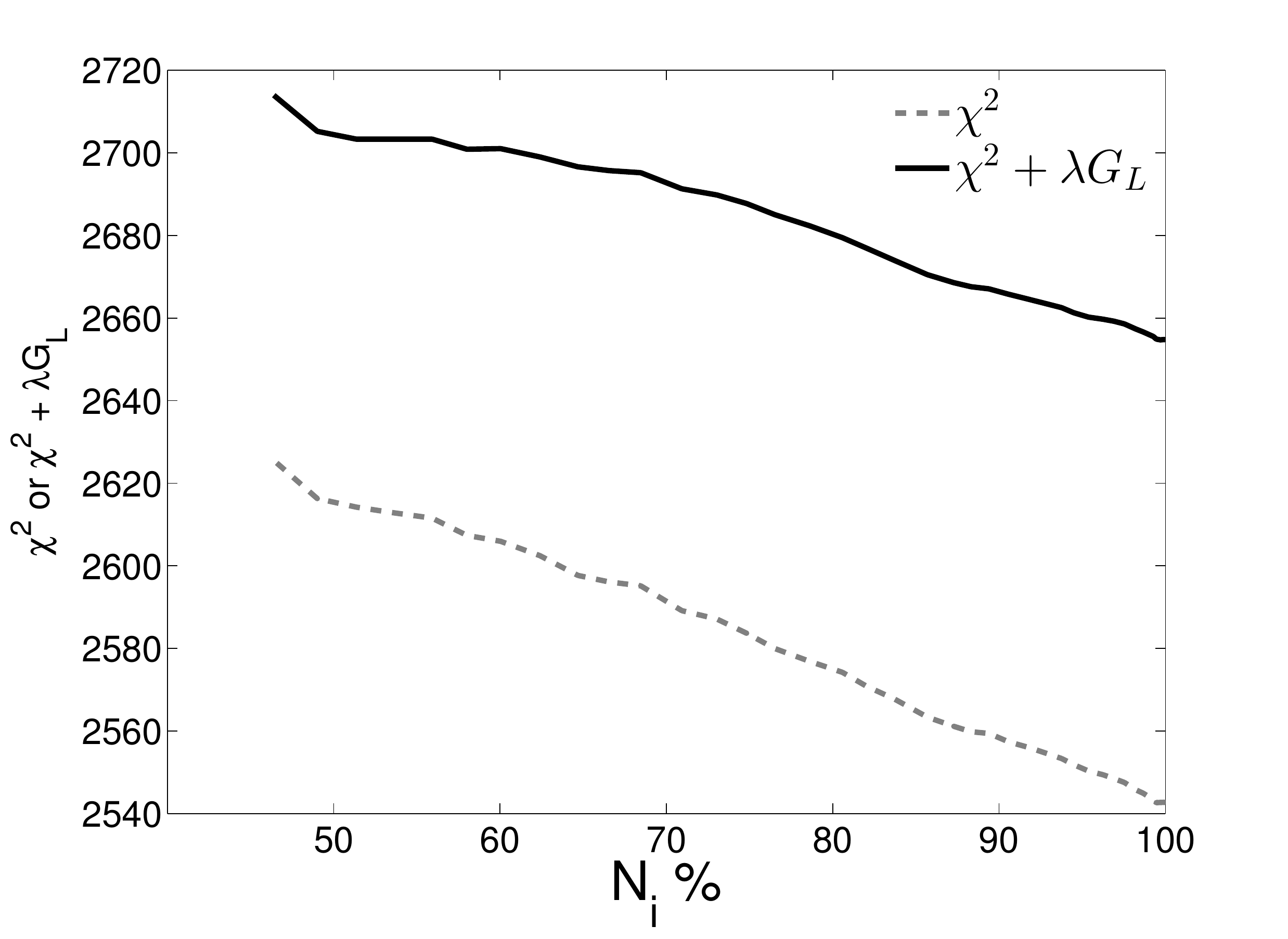}
\caption{Dependence of the image residuals $\chi^2$ (thick black) and overall merit function $\chi^2 + \lambda G_L$ (dashed grey) on the number of traced image pixels $N_i$ where the evidence term of equation (4) is not used and $\lambda$ is fixed. A reduction in $N_i$ results in a higher $\chi^2$ and higher overall merit function $\chi^2 + \lambda G_L$}
\label{figure:NiEviOff}
\end{figure}

We note that we calculate the NDOF as the number of image pixels which successfully trace within the circular source plane aperture minus the total number of source plane pixels, which is fixed to 400. As discussed in \cite{Suyu2006}, regularization correlates source pixels thus decreasing the number of effective source pixels and increasing the 'true' NDOF. Since in the example presented here we allow the regularization weight to change, the number of effective source pixels is varying and thus this does contribute to a change in the 'true' NDOF. When referring to the NDOF we are ignoring correlations between source pixels due to regularization, instead holding its value fixed to 400. Either way, our interpretation of the $N_i$ bias does not depend on our definition of the NDOF, as figures \ref{figure:NiRegLH}, \ref{figure:NiChiReg} and \ref{figure:NiEviOff} are plotted and discussed as functions of $N_i$.

We first reconstruct image 1 using the input lens model and investigate how $\lambda$ and $\ln\epsilon$ scale with $N_i$ by changing the source plane aperture radius. The $\chi^2$ term in the evidence is calculated in the same fixed image mask for each aperture radius. Figure \ref{figure:NiRings} shows the range of mask radii we use. 

The results plotted in figure \ref{figure:NiRegLH} show that as $N_i$ increases, the regularization weight (selected by maximizing the evidence) also increases. However, the evidence falls with increasing $N_i$ and thus lens optimization's which allow the number of image pixels that trace to the source plane to vary are biased towards lower magnification parameter sets. This is a manifestation of the $\sigma-q-\alpha$ degeneracy.

To understand this behaviour at a more fundamental level, we investigate this further in figure \ref{figure:NiChiReg} by plotting the variation of two components of the evidence as expressed in equation (\ref{eqn:evidence}). The first component is simply $\chi^2$, which as expected, reflects poorer image reconstruction by becoming larger as $N_i$ is reduced. The second component we plot is the sum of the second, third and fourth terms in equation (\ref{eqn:evidence}) which together account for the regularization dependence of the evidence. The figure shows that despite poorer image reconstruction, the overall behavior of the evidence is dominated by the more rapidly varying regularization terms (note that the summation of both terms determines the behavior of $-2 \ln \epsilon$ since the last term in equation (\ref{eqn:evidence}) is constant).

Since the evidence appears to cause a bias towards solutions with fewer traced image pixels, we repeat this analysis with a fixed value of $\lambda$ and the non-Bayesian merit function $G = \chi^2 + \lambda G_{L}$, originally advocated by WD03. The results are plotted in figure \ref{figure:NiEviOff} which shows that, in exactly the same way that $\chi^2$ in figure \ref{figure:NiChiReg} behaves, the bias is now present in the opposite sense, whereby solutions favored are those with the highest value of $N_i$. This is more intuitive given that image reconstruction should be more accurate when given every possible data point. This demonstrates that regardless of whether the evidence is used or not, lens modeling with a varying NDOF will be biased to solutions which either minimize or maximize $N_i$ and accurate lens modeling with the SPLE profile therefore requires this to be fixed throughout model optimization.

In the case of the SIE profile, while this bias is still present, the $\sigma-q-\alpha$ degeneracy, which allows the source to geometrically scale, is not. Therefore, even though square SLI allows the NDOF to vary, only parameter sets near the true SIE model actually give accurate image reconstructions. The variation of the $\chi^2$ term in the evidence for the SIE profile dominates the variation of the regularization terms. In the case of the SPLE profile, the evidence coupled with the $\sigma-q-\alpha$ degeneracy tries to minimize $N_i$ which corresponds to positively biasing $\alpha$ thus explaining the incorrect lens models initially found. Of course a cut-off will be reached when the expanded source covers the entire source plane and the increase in the contribution from the regularization terms in the evidence from reducing $N_i$ is offset by the more rapidly increasing $\chi^2$ term as source light is lost and the observed lensed image becomes poorly fit.

\begin{figure}
\centering
\includegraphics[width=0.5\textwidth]{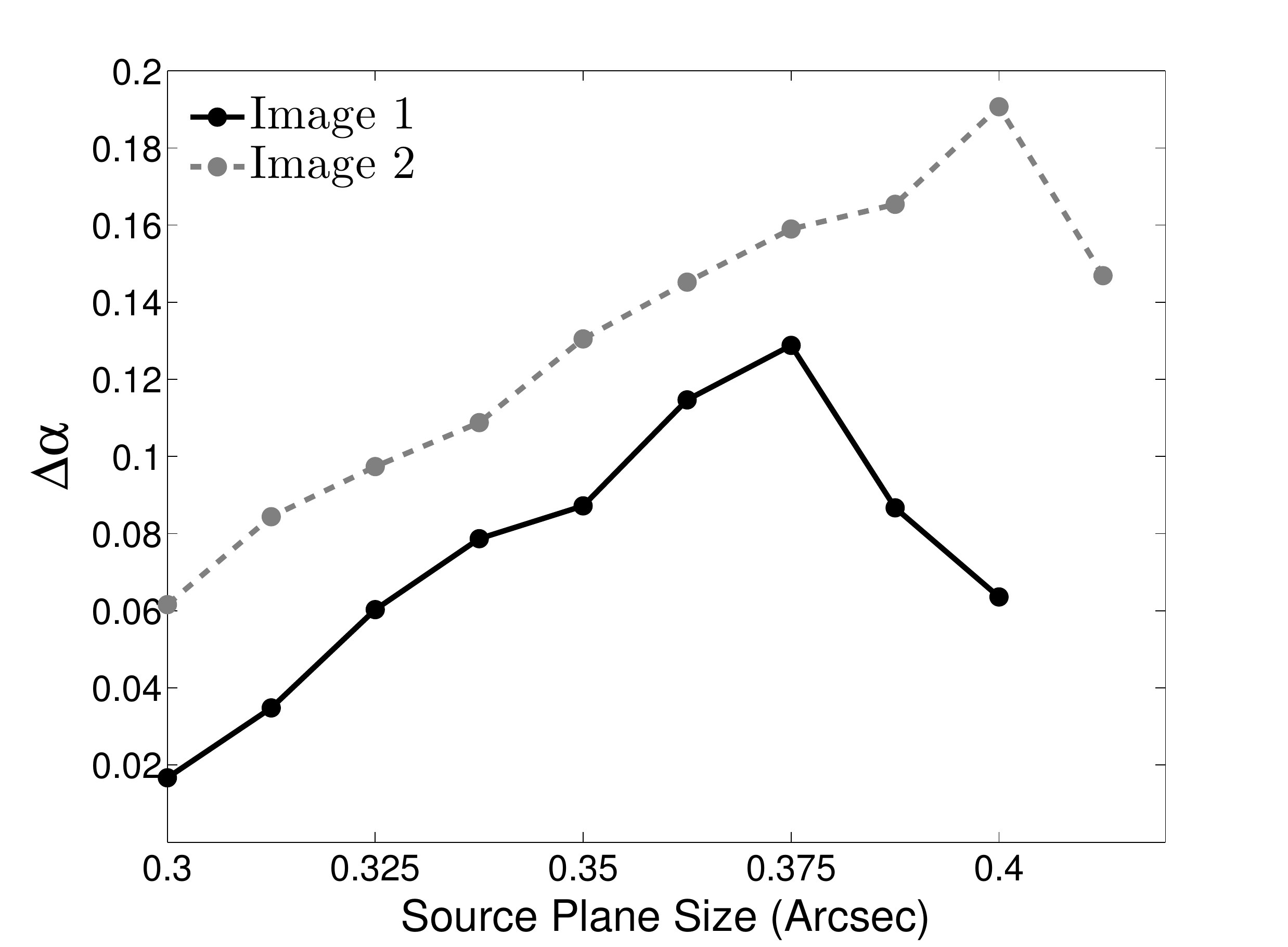}
\caption{$\alpha$ bias, $\Delta\alpha = \alpha_{\rm calc} - \alpha_{\rm true}$, plotted against source plane size for both images 1 and 2. All points are generated using a full {\tt MultiNest} nonlinear search.}
\label{figure:AlSorBias}
\end{figure}

One would therefore expect a relationship between the source plane size and the bias in $\alpha$, $\Delta\alpha = \alpha_{\rm calc} - \alpha_{\rm true}$. Reverting back to the Bayesian merit function, a larger source plane allows solutions corresponding to higher $\alpha$ to be attained before the source expands outside the source plane boundary. Figure \ref{figure:AlSorBias} confirms that $\Delta\alpha$ does increase with increasing source plane size as expected. This continues until a turnover point when the source plane starts to become larger than the maximum extent of the source expansion allowed by the $\sigma-q-\alpha$ degeneracy. Just beyond the turnover point are intermediate values where $N_i$ still varies but the source plane becomes large enough to lessen the bias. The turnover occurs at different source plane sizes between images 1 and 2 due to differences in the source position with respect to the image caustic and thus differences in $N_i$.

The requirement that the NDOF must be fixed during lens optimization is naturally satisfied by the adaptive SLI method since every source plane pixelization is derived directly from the magnification map and is independent of the source plane size which keeps $N_i$ fixed. Other inversion methods found in the literature have also identified this problem \citep[for example,][]{Vegetti2009,Suyu2013,Collett2014}, although the level of efficacy with which it has been managed is somewhat variable. Nevertheless, as we discuss below, there are other more pertinent biases which have not been fully appreciated and which are explicitly addressed by adaptive SLI.

\subsubsection{Bias 2: Discreteness Biases}\label{GridAlign}

Following our discussion in the previous sub-section, we fix the NDOF hereafter when using square SLI, by increasing the source plane size to $0.7'' \times 0.7''$. We also narrow the priors on the lens offsets to $\pm0.02''$ and on $\phi$ to $\pm 2 \,^{\circ}$ to reduce caustic movement and we use a prior on $\alpha$ of $\pm0.15$ to limit the largest caustic size. These changes are applied to the inversion of both images 1 and 2 and we maintain a source plane resolution of $20 \times 20$ pixels. As discussed, these changes to the inversion setup lead to a far less efficient and less robust inversion, demonstrating the disadvantage of using a fixed grid which is initialized independently of the lens model. Results presented using adaptive SLI retain the wider priors given at the beginning of this section, i.e., those used for the SIE profile and $\alpha$ between 1.75 and 2.25. 

Despite these changes, SPLE modeling with square SLI continues to be both inaccurate and dependent on the way in which the source plane is pixelized. Similar findings have also been made by \citet{Suyu2013} who found that the dominant systematic uncertainty in their lens modeling is source plane resolution and \citet{Tagore2014} who showed that the results of a SIE lens model change if the setup of either the observed image (noise map, telescope pointing) or source plane (regularization scheme) is varied. Clearly, these systematic effects must be eradicated in order to robustly determine lens model parameters and their error distributions.

The importance of data discretization was, in fact, already highlighted in figure \ref{figure:LHNoise}, where adaptive SLI's noisy parameter space is a direct consequence of changing just the source plane discretization. The figure shows that a change in the data discretization can give a relatively large change in the value of $\ln \epsilon \simeq 10$. The scale of this change is comparable to the scale over which solutions within the $\sigma-q-\alpha$ degeneracy vary and therefore selecting a single source plane discretization with square SLI will generally give rise to a significantly biased solution. We hereafter refer to this effect as `discreteness bias'.

We test for discreteness biases by calculating lens models for images 1 and 2 with square SLI, using the setup described above to fix the NDOF. However, for each lens model, we shift the position of the square grid by a fractional pixel width. This gives a small change in the data discretization such that there is a minor shift in the overall allocation between image and source pixels for an identical lens model. For each image we perform modeling using nine different phase shifts spread over a $3 \times 3$ pattern, with sub-gridding off and with sub-gridding of $4 \times 4$. Every nonlinear search uses 200 active {\tt MultiNest} points. 

The results are shown in table \ref{table:PhaseTable}. The third and fourth columns show the value of $\alpha$ obtained for each image without sub-gridding. Phase shifting of the square grid directly impacts the value of $\alpha$ estimated, with several of the values of $\alpha$ being significantly biased for both images. This demonstrates the effect of discreteness bias. For comparison, in the last row of the table we also show the results of adaptive SLI using 200 clusters, an arbitrarily large source plane and 300 active {\tt MultiNest} points. As the table shows, adaptive SLI also calculates an incorrect lens model without sub-gridding because pixel aliasing effects in the image dominate the inversion. Like \citet{Tagore2014}, we also find that changing the synthetic image noise realization changes the resulting set of most probable lens parameters obtained if image sub-gridding is not used. 

The results of removing pixel aliasing by applying image sub-gridding are shown in the fifth and sixth columns of table \ref{table:PhaseTable}. The values of $\alpha$ obtained from image 1 with square SLI are now consistent with the input lens model, although an element of scatter still indicates the effect of source plane discretization. However, $\alpha$ obtained with square SLI from image 2 continues to be significantly discrepant with the input value, with many values inconsistent at the $3\sigma$ confidence level. Applying sub-gridding with adaptive SLI returns accurate values of $\alpha$ for both images.

\begin{figure*}
\centering
\includegraphics[width=0.97\textwidth]{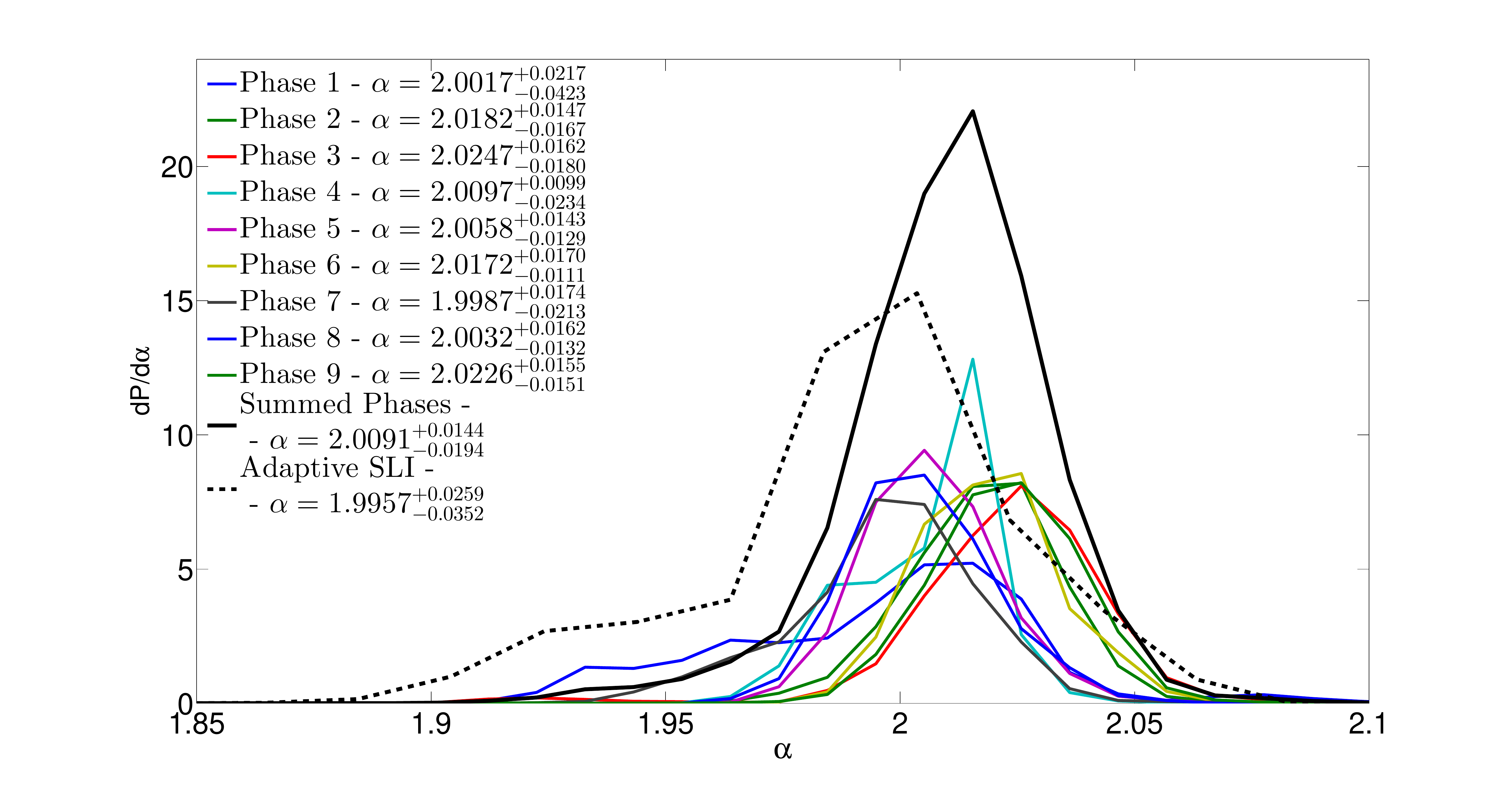}
\includegraphics[width=0.97\textwidth]{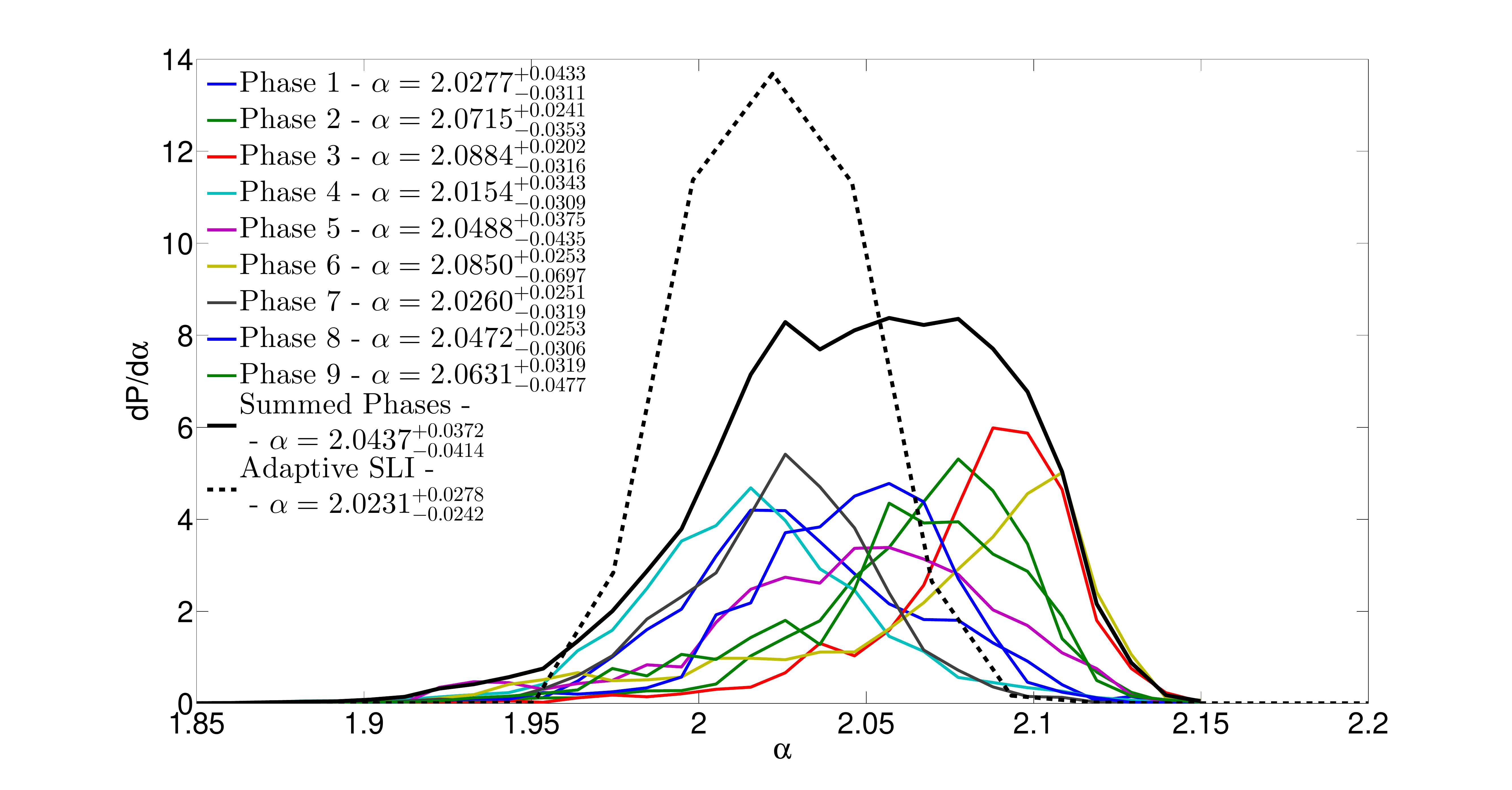}
\caption{Posterior distribution function (PDF) of $\alpha$ given by square SLI, using nine phase shifted grids on image 1 (top) and image 2 (bottom). Results correspond to the third and fourth columns of table \ref{table:PhaseTable} (see caption for details). The PDF of individual phase shifts is scaled by $\frac{1}{3}$ for clarity. The figure demonstrates the variation between PDFs of different phase shifts resulting from discretization biases. The thick black line is the sum of the nine phase shifted PDFs and shows consistency with the lens model using image 1 but not image 2. The dashed line shows the PDF calculated using adaptive SLI, set up with 200 clusters and 300 {\tt MultiNest} live points for both images.} 
\label{figure:GridAlign_isub4}
\end{figure*}

We represent these results graphically in figure \ref{figure:GridAlign_isub4}, where the marginalized one-dimensional posterior distribution function (PDF) of the sum of all nine phase shifts for square SLI, marginalized over $\alpha$, is plotted for both images with a thick black line. Alongside this, the PDF of each individual phase shift is also shown, scaled by $\frac{1}{3}$ for clarity. The PDF given by adaptive SLI is shown with a dashed black line. The figure shows that although the summed PDF for image 1 is consistent with the input value of 2.0, in image 2, this is much less so. Since narrower lens model priors were introduced for square SLI in this section, the PDF for $\alpha$ for image 2 falls off more rapidly towards $\alpha=2.15$ which artificially lessens the inconsistency that would have otherwise been observed. Weighting the calculation of the summed PDF by evidence gives near-identical results owing to the small difference between the evidence values of each individual PDF.

\begin{figure*}
\centering
\includegraphics[width=0.97\textwidth]{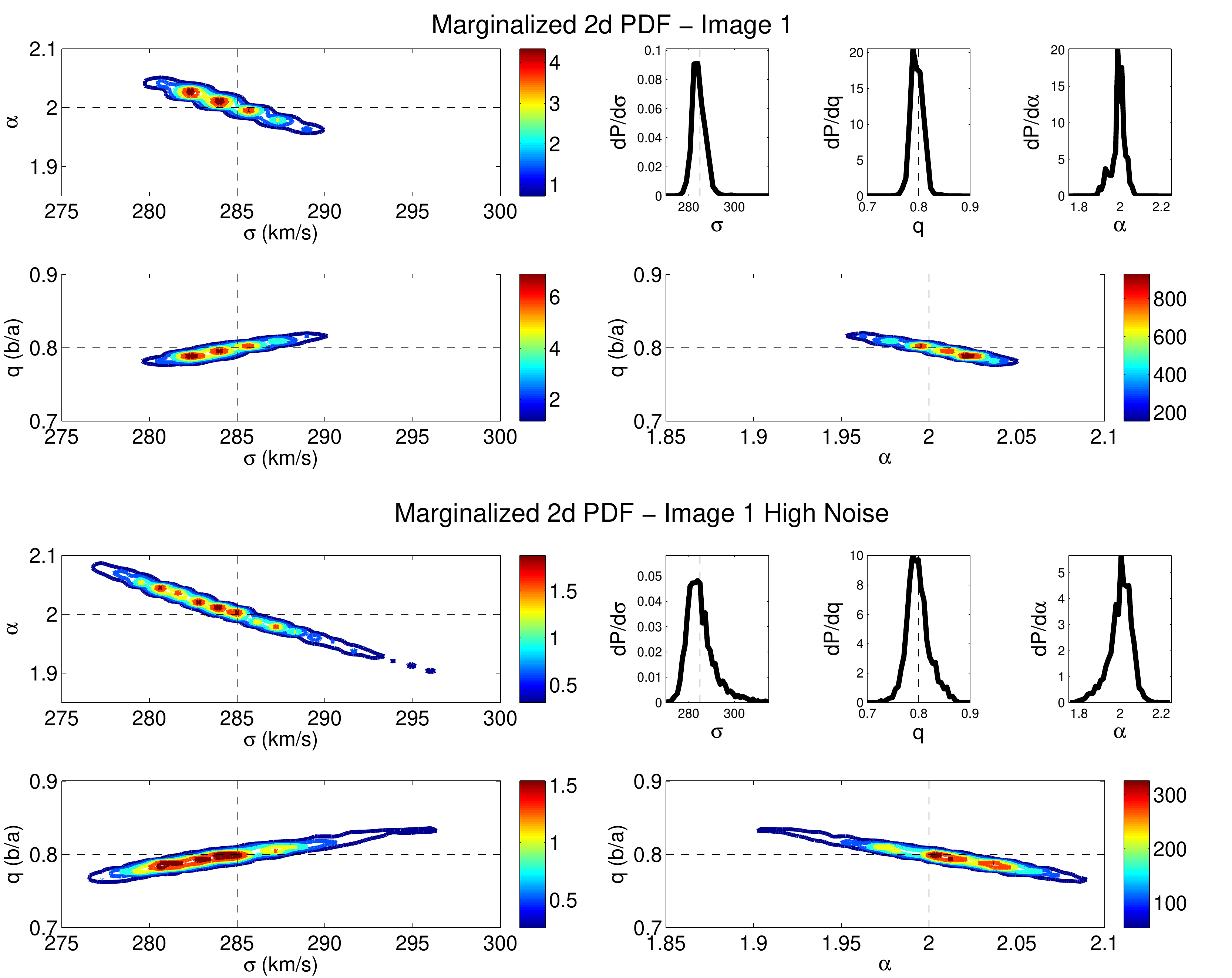}
\caption{Two dimensional PDFs of $\sigma$, $q$ and $\alpha$ given by adaptive SLI for image 1 (top) and a lower S/N version of image 1 (bottom). The one dimensional PDFs for these parameters are shown in the top right corner for each image.} 
\label{figure:Cont1}
\end{figure*}

\begin{figure*}
\centering
\includegraphics[width=0.97\textwidth]{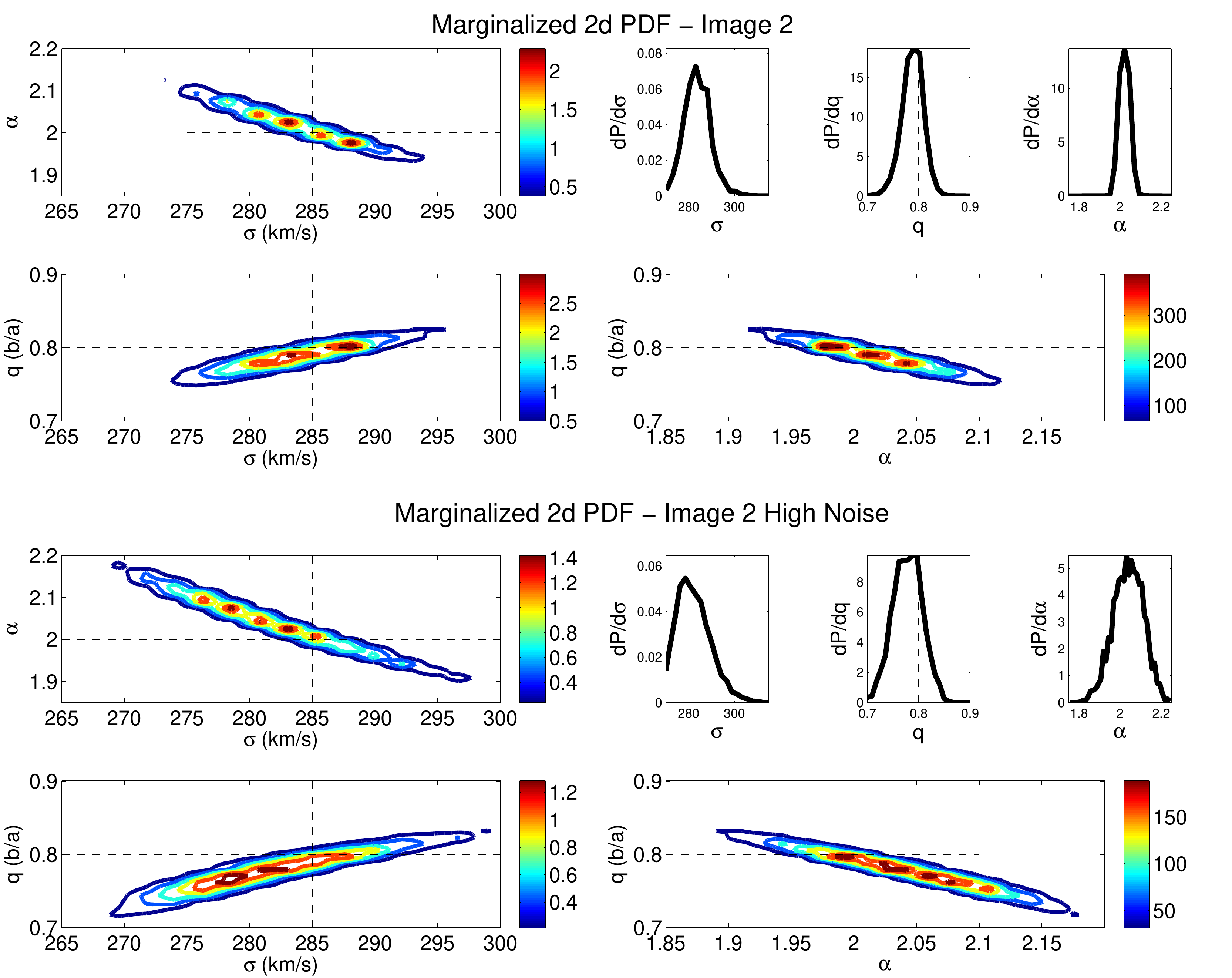}
\caption{Two dimensional PDFs of $\sigma$, $q$ and $\alpha$ given by adaptive SLI for image 2 (top) and a lower S/N version of image 2 (bottom). The one dimensional PDFs for these parameters are shown in the top right corner for each image.} 
\label{figure:Cont2}
\end{figure*}

Conversely, as table \ref{table:PhaseTable} and figure \ref{figure:GridAlign_isub4} show, adaptive SLI is accurate for both images. The fact that data discretization is different and unique for every trial lens model parameter set, regardless of the size of the change in parameters, underlies the reliability of the method (see appendix \ref{AppA} for a more detailed discussion). In addition, figures \ref{figure:Cont1} and \ref{figure:Cont2} show the two-parameter adaptive SLI PDFs for images 1 and 2 respectively. In each of these figures, we also show the results of a higher noise run where the S/N was lowered to 70 in each image. The figures show the $\sigma - q - \alpha$ degeneracy previously discussed and also that parameter errors increase for poorer quality data. This effect is much less obvious with square SLI due to the inherent discreteness biases present. Furthermore, in appendix \ref{AppB}, we demonstrate that adaptive SLI again accurately models a SPLE for both images 1 and 2 for a variety of source plane resolutions, levels of image subgridding and inversion setups.

Figure \ref{figure:GridAlign_isub4} shows the PDF of individual square SLI runs are generally both narrower and more sharply peaked than the PDF given either by their summation or adaptive SLI. Discreteness biases lead to higher evidence values being calculated at the favored lens model, resulting in significant error under estimation. While this can be alleviated by the summing of multiple inversions with differing methods of data discretization, as was done in \cite{Suyu2013} and figure \ref{figure:GridAlign_isub4}, this still only provides an approximation of the errors and as shown for image 2 may still ultimately give a biased lens model. In appendix \ref{AppA} we find error underestimation occurs if we fix adaptive SLI's initialization, thus reintroducing discreteness biases. Our general conclusion is that without fully accounting for all systematics associated with data discretization, a comprehensive estimation of all the lens models associated errors is not possible. Moreover, by accounting for these systematics adaptive SLI permits a more accurate calculation of the marginalized evidence improving the prospects for accurate model comparison. 

In addition to the effect of lowering S/N, we also investigate the effect of changing source plane resolution. We use the same phase shifting methodology as previously but now use a higher source plane resolution of $36 \times 36$ pixels with square SLI. In this case, to improve efficiency, we calculate $\lambda$ by maximizing equation (\ref{eqn:evidence}) for each phase using the input lens model and then keep $\lambda$ fixed to that value for the entire inversion. We then minimize the basic merit function $G = \chi^2 + \lambda G_L$ instead of the evidence. Of course this approach can not be used for real observations although we discuss strategies for improving efficiency with real data in section \ref{NpixBias}. 

\begin{figure*}
\centering
\includegraphics[width=0.97\textwidth]{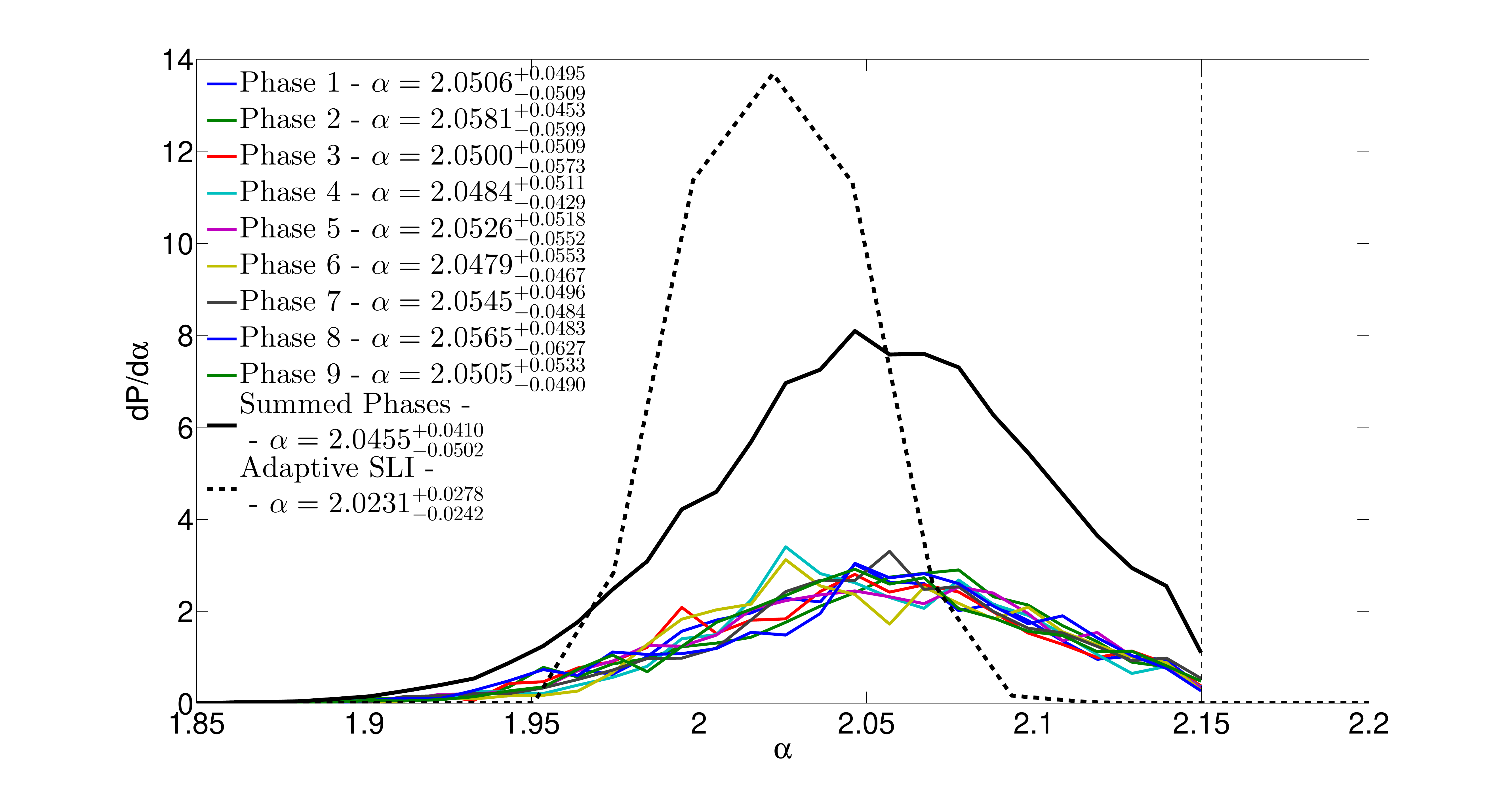}
\caption{As figure \ref{figure:GridAlign_isub4} but showing the PDF of $\alpha$ given by square SLI using 9 high resolution $(36 \times 36)$ phase shifted grids on image 2. The Bayesian wrapper of S06 is used to initially set $\lambda$ but is then switched off for the inversion. Results correspond to the seventh column of table \ref{table:PhaseTable}. The PDF using adaptive SLI is that given in the bottom panel of figure \ref{figure:GridAlign_isub4}. The figure shows that a higher resolution source plane does not remove discretization biases.} 
\label{figure:GridAlign_sp36}
\end{figure*}

Figure \ref{figure:GridAlign_sp36} shows the resulting PDFs obtained with the higher resolution reconstruction. The most striking feature is that they show a much broader PDF. This is a consequence of removing the Bayesian wrapper and fixing the regularization weight. The variation in the PDFs for each source grid phase indicates that discreteness biases are still present, however, the presence of an additional bias discussed below in section \ref{NpixBias} leads to greater consistency amongst their overall $\alpha$ estimation. A cut off at the upper $\alpha$ prior is also seen, showing the disadvantage of being forced to narrow priors to ensure the NDOF remains fixed. It is clear from this test that discretization biases remain present regardless of the change in source plane resolution.

\begin{table*}
\begin{tabular}{ l | l | l | l | l | l | l } 
\multicolumn{1}{p{0.9cm}|}{\centering Phase shift x} 
& \multicolumn{1}{p{0.9cm}|}{\centering Phase shift y} 
& \multicolumn{1}{p{2.5cm}|}{\centering Image 1 \\ No sub-gridding} 
& \multicolumn{1}{p{2.5cm}|}{\centering Image 2 \\ No sub-gridding} 
& \multicolumn{1}{p{2.5cm}|}{\centering Image 1 \\ Sub-gridding $4 \times 4$}
& \multicolumn{1}{p{2.5cm}|}{\centering Image 2 \\ Sub-gridding $4 \times 4$}
& \multicolumn{1}{p{2.5cm}}{\centering Image 2 \\ High resolution ($36 \times 36$)}
\\ \hline
& & & & & & \\[-4pt]
0. & 0. & $1.9533 ^{+0.0071} _{-0.0058}$ & $2.0616 ^{+0.0165} _{-0.0213}$ & $2.0017 ^{+0.0217} _{-0.0423}$ & $2.0277 ^{+0.0433} _{-0.0311}$ & $2.0506 ^{+0.0495} _{-0.0509}$ \\[2pt]
0.25 & 0. & $1.9944 ^{+0.1060} _{-0.0345}$ & $1.9480 ^{+0.1064} _{-0.0379}$ & $2.0182 ^{+0.0147} _{-0.0167}$ & $2.0715 ^{+0.0241} _{-0.0353}$ & $2.0581 ^{+0.0453} _{-0.0599}$ \\[2pt]
0.5 & 0. & $2.0631 ^{+0.0171} _{-0.0159}$ & $1.9777 ^{+0.0087} _{-0.0128}$ & $2.0247 ^{+0.0162} _{-0.0180}$ & $2.0884 ^{+0.0202} _{-0.0316}$ & $2.0500 ^{+0.0509} _{-0.0573}$ \\[2pt]
0. & 0.25 & $2.0368 ^{+0.0117} _{-0.0153}$ & $2.0584 ^{+0.0118} _{-0.0156}$ & $2.0097 ^{+0.0099} _{-0.0234}$ & $2.0154 ^{+0.0343} _{-0.0309}$ & $2.0484 ^{+0.0511} _{-0.0429}$ \\[2pt]
0.25 & 0.25 & $2.0238 ^{+0.0165} _{-0.0191}$ & $1.9581 ^{+0.0114} _{-0.0159}$ & $2.0058 ^{+0.0143} _{-0.0129}$ & $2.0488 ^{+0.0375} _{-0.0435}$ & $2.0526 ^{+0.0518} _{-0.0552}$ \\[2pt]
0.5 & 0.25 & $2.0326 ^{+0.0150} _{-0.0074}$ & $1.9860 ^{+0.0066} _{-0.0123}$ & $2.0172 ^{+0.0170} _{-0.0111}$ & $2.0850 ^{+0.0253} _{-0.0697}$ & $2.0479 ^{+0.0553} _{-0.0467}$ \\[2pt]
0. & 0.5 & $2.0415 ^{+0.0120} _{-0.0156}$ & $2.0871 ^{+0.0095} _{-0.0115}$ & $1.9987 ^{+0.0174} _{-0.0213}$ & $2.0260 ^{+0.0251} _{-0.0319}$ & $2.0545 ^{+0.0496} _{-0.0484}$ \\[2pt]
0.25 & 0.5 & $2.0137 ^{+0.0357} _{-0.0189}$ & $2.0905 ^{+0.0070} _{-0.1032}$ & $2.0032 ^{+0.0162} _{-0.0132}$ & $2.0472 ^{+0.0253} _{-0.0306}$ & $2.0565 ^{+0.0483} _{-0.0627}$ \\[2pt]
0.5 & 0.5 & $2.0319 ^{+0.0174} _{-0.0074}$ & $2.0047 ^{+0.0066} _{-0.0223}$ & $2.0226 ^{+0.0155} _{-0.0151}$ & $2.0631 ^{+0.0319} _{-0.0477}$ & $2.0505 ^{+0.0533} _{-0.0490}$ \\[4pt]
\hline
\multicolumn{2}{p{1.8cm}|}{ } & & & & & \\[-4pt]
\multicolumn{2}{p{1.8cm}|}{\centering Average} & $2.0289 ^{+0.0187} _{-0.0635}$ & $1.9921 ^{+0.0779} _{-0.0365}$ & $2.0091 ^{+0.0144} _{-0.0194}$ & $2.0437 ^{+0.0372} _{-0.0414}$ & $2.0455 ^{+0.0410} _{-0.0502}$ \\[2pt]
\hline
\multicolumn{2}{p{1.8cm}|}{ } & & & & & \\[-4pt]
\multicolumn{2}{p{1.8cm}|}{Adaptive SLI} & $1.9481 ^{+0.0072} _{-0.0038}$ & $2.0934 ^{+0.0031} _{-0.0077}$ & $1.9958 ^{+0.0258} _{-0.0354}$ & $2.0231 ^{+0.0278} _{-0.0242}$ & N/A
\end{tabular}
\caption{The values of $\alpha$ estimated using square SLI on a set of nine phase shifted grids. Each result corresponds to an individual non-linear search with the NDOF fixed for every lens model. The third to sixth columns the source plane is size $0.7'' \times 0.7''$ and resolution $20 \times 20$ pixels. The first third and fourth columns are results for images 1 and 2 without image sub-gridding and the fifth and sixth apply to sub-gridding of $4 \times 4$. The seventh column corresponds to a high resolution, $36 \times 36$, square grid for image 2 with sub-gridding of $4 \times 4$ and fixed $\lambda$ set by maximizing equation (\ref{eqn:evidence}) and merit function $G = \chi^2 + \lambda G_L$. The bottom two rows show the marginalized value of $\alpha$ when summed over all 9 phase shifts and the value given by adaptive SLI. The PDFs are plotted in figures \ref{figure:GridAlign_isub4} and \ref{figure:GridAlign_sp36}. }
\label{table:PhaseTable}
\end{table*}

\subsubsection{Bias 3: Effective Source Resolution}\label{NpixBias}

The third and final bias is again relevant for the SPLE model and also more generally, those with a degeneracy that allows geometric source scaling. The degeneracy arises with any fixed source plane pixelization if the regularization weight is not correctly optimized by finding the maximum evidence with each iteration, or if a fixed regularization weight is used.

To demonstrate this effect, we use square SLI with image 1 and a SPLE lens model fixed with the input set of parameters. We vary the source plane resolution in steps from $12 \times 12$ to $28 \times 28$ and reconstruct the source with each resolution, keeping the source plane size fixed. With a fixed source plane size and fixed lens model, the number of image pixels traced to the source plane also remains fixed and hence this test is not sensitive to bias 1 where $N_i$ varies. We investigate how the figure of merit $\chi^2 + \lambda G_L$ varies with source plane resolution when $\lambda$ is fixed at the optimal value for a $20 \times 20$ pixel source plane and then how the full evidence varies when $\lambda$ is optimized for each resolution.

\begin{figure}
\centering
\includegraphics[width=0.48\textwidth]{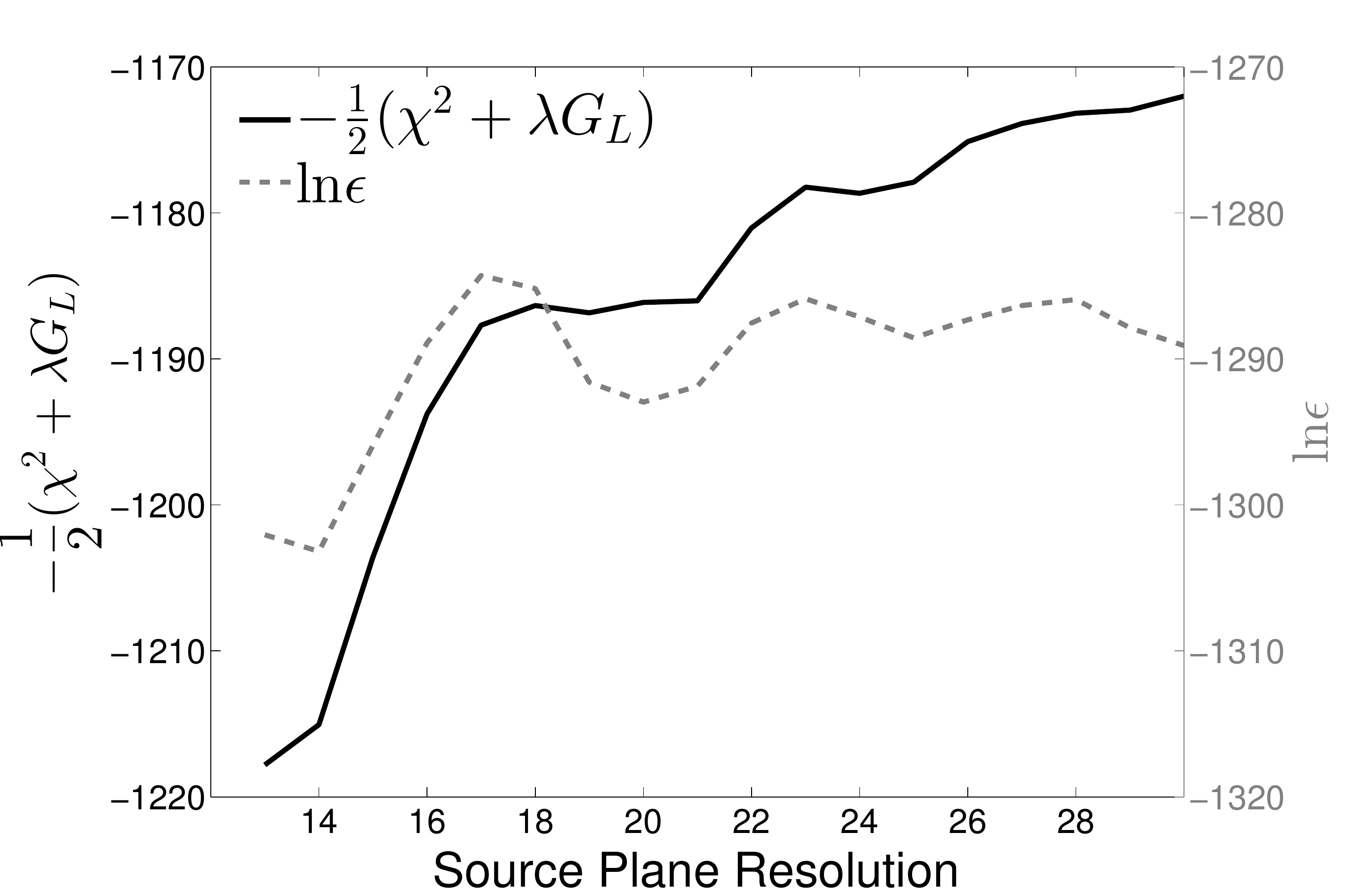}
\caption{Variation of $\ln \epsilon$ and $-(\chi^2+\lambda G_L)/2$ with source plane resolution for square SLI. In the case of $\ln \epsilon$, $\lambda$ is optimized for each source plane resolution by maximizing the evidence. For the figure of merit $-(\chi^2+\lambda G_L)/2$, $\lambda$ is fixed at the optimal value for a source plane resolution of 20. The source plane resolution corresponds to the number of pixels along one edge of a square grid of pixels for a fixed source plane size of $0.7'' \times 0.7''$.}
\label{figure:SrcResLH}
\end{figure}

Figure \ref{figure:SrcResLH} shows the results. When $\lambda$ is fixed, the figure of merit improves near-monotonically to higher source plane resolutions. When $\lambda$ is optimized, the evidence remains more constant (modulo the variation due to data discretization effects previously discussed). This shows that fixing $\lambda$ biases lens models with degeneracies which allow the source to geometrically scale on a fixed resolution grid. The reason for this is because the number of source pixels representing the significant source flux varies, thus mimicking a changing source resolution. We illustrate this in figure \ref{figure:SPLE} for the SPLE model where we outline in bold those source plane pixels that have a flux that is a factor of 1.5 above the read noise. The total number of these pixels is labeled by $N_{\rm Pix}$ in the figure. $N_{\rm Pix}$ can be considered a measure of the effective source plane resolution for reconstruction. In this way, the figure shows that square SLI has a varying effective source plane resolution with different SPLE lens model parameterizations within the $\sigma - q - \alpha$ degeneracy. 

The expectation with the SPLE model is therefore that fixed source pixelizations with non-optimized regularization weights bias lens model parameters to high values of $\alpha$ where the effective source resolution is increased. This behavior is clearly seen in figure \ref{figure:GridAlign_sp36} for fixed $\lambda$ where the PDFs for each source grid phase shift unanimously agree on a value of $\alpha$ that is higher than the input value. We note that the behavior is also seen to a lesser extent in figure \ref{figure:GridAlign_isub4} where $\lambda$ is optimized and yet there remains a bias in the summed PDF for $\alpha$. In appendix \ref{AppA} we discuss this result more, suggesting it may infact be associated with discreteness biases.

The bottom row of figure \ref{figure:SPLE} shows that by adapting to the magnification, adaptive SLI retains a more equal effective source resolution for all lens models. While $N_{\rm Pix}$ still varies, it fluctuates about a mean value and ultimately the random nature of the pixelization between iterations of the lens modeling average away any resulting systematics, in the same manner in which discretization biases are removed. Source pixelization schemes suggested in the literature which adapt to the lens model by scaling the size of the source plane according to the caustic size (e.g. \cite{Suyu2006, Collett2014}) will remove the effective resolution bias to first order, but it is possible that the bias is not completely removed. This is most likely to occur when the caustic is non-symmetric, for example, in the presence of external shear, however this is something we have not investigated in the present work.

\subsection{Lens Model With Fixed Regularization}\label{FixReg}

We have shown how adaptive SLI, when optimizing $\lambda$ with each lens model iteration, removes biases that occurs with the SPLE model when using a fixed source plane pixelization. What we have not considered is how well adaptive SLI copes if $\lambda$ is not optimized with every lens model iteration. A reduction in the rate of the number of $\lambda$ optimization's per lens model iteration has the advantage of a significant increase in modeling efficiency, since this reduces the number of times the computationally expensive determinants in equation (\ref{eqn:evidence}) must be evaluated.

To test the feasibility of adaptive SLI with such a reduced rate of $\lambda$ optimization, we carry out a simple test where we use adaptive SLI with a fixed $\lambda$ and the same inversion setup as the previous section. For both images 1 and 2 we perform lens modeling three times, each using a value of $\lambda$ of 0.5, 1.0 and 1.5 times the optimal value found by maximizing the evidence.

The results are shown in table \ref{table:Npix_EviOff}. All values of $\alpha$ are consistent with the input lens model and minimal scatter is seen between results using different $\lambda$. From a modeling efficiency point of view, this is very encouraging, demonstrating that a reduced rate of optimizing $\lambda$ is feasible.

Both \cite{Vegetti2009, Collett2014} employ this strategy, although given real data is used their initial lens model is estimated using a fixed $\lambda$ corresponding to over regularization. This model then maximizes equation (\ref{eqn:evidence}), giving a new $\lambda$ which is held fixed to estimate the final lens model. This process therefore performs two non-linear searches with the Bayesian wrapper off, which is only used once to optimize $\lambda$ after the initial run. While this strategy is clearly worthy of future investigation for adaptive SLI, its handling of data discretization means it performs lens modeling at comparatively lower source plane resolutions anyway, for which the optimization of $\lambda$ for every set of lens parameters remains fast to compute and therefore viable. While this should be generally be preferred when possible, provided the aforementioned approximations are not dominant, a hybrid of both strategies may be best when the analysis of large data sets or high resolution images is considered. This will be of major consideration in the development of adaptive SLI into a streamlined strong lens analysis pipeline.

\begin{table}
\begin{tabular}{ l | l | l }
\multicolumn{1}{p{0.8cm}|}{\centering Image} 
& \multicolumn{1}{p{0.6cm}|}{\centering $\lambda$}
& \multicolumn{1}{p{1.5cm}}{\centering $\alpha$}
\\ \hline
& & \\[-6pt]
1 & 1.5 $\lambda_{\rm opt}$ & $1.9990 ^{+0.0355} _{-0.0331}$ \\[1pt]
1 & 1.0 $\lambda_{\rm opt}$ & $1.9909 ^{+0.0366} _{-0.0321}$ \\[1pt]
1 & 0.5 $\lambda_{\rm opt}$ & $1.9724 ^{+0.0384} _{-0.0364}$ \\[1pt] \hline
& & \\[-9pt]
2 & 1.5 $\lambda_{\rm opt}$ & $2.0211 ^{+0.0156} _{-0.0252}$ \\[1pt]
2 & 1.0 $\lambda_{\rm opt}$ & $2.0225 ^{+0.0297} _{-0.0269}$  \\[1pt]
2 & 0.5 $\lambda_{\rm opt}$ & $2.0151 ^{+0.0202} _{-0.2900}$ \\

\end{tabular}

\caption{Marginalized $\alpha$ for images 1 and 2 using the merit function $G = \chi^2 + \lambda G_L$ with adaptive SLI. $\lambda$ is fixed to 0.5, 1.0 and 1.5 times the optimal value, $\lambda_{\rm opt}$, set by maximizing the evidence.}

\label{table:Npix_EviOff}

\end{table}

\section{Summary And Discussion}\label{Discussion}

We have presented adaptive semi-linear inversion (SLI), a new method for the inversion of gravitationally lensed extended sources. The source plane pixelization is determined by clustering the coordinates of all traced image pixels with an h-means algorithm, deriving a source plane pixelization which matches the lens magnification for every lens model parameter set. The distinguishing feature of adaptive SLI is that it does this using a random initialization and therefore the discretization of source plane data is handled in a completely different, unique way for every lens model. The method then efficiently samples the underlying posterior distribution of degenerate lens models while naturally accounting for systematics which otherwise bias lens modeling. We have demonstrated this unique feature using the specific example of a SPLE lens model.

In this paper, we have compared adaptive SLI with the standard semi-linear inversion method of WD03 which uses a fixed square grid of pixels to discretize the source plane. In this comparison, we have used two realistic simulated images to highlight the benefits of a source plane pixelization which adapts to the lens magnification. Our selection of the standard SLI method of WD03 for comparison with adaptive SLI was for simplicity, but many of the consequences arising from use of a square grid apply to use of fixed source pixelizations generally.

We have discussed two key biases inherent to pixelized inversions and we have demonstrated how adaptive SLI removes them. These are:

\begin{itemize}

\item[(i)] Dependent on the figure of merit chosen for optimization of the lens model parameters, methods which allow the number of degrees of freedom (NDOF) to vary between model iterations, due to image pixels tracing outside the source plane, try to either maximize or minimize the NDOF. Such extremes in the NDOF are achieved by lens model parameters which lie a significant distance from the correct parameters. Although a fixed NDOF can be ensured with any pixelized inversion method, this typically results in a less efficient and robust inversion as well as the requirement that priors on lens parameters are narrowed. The pixelizations calculated by adaptive SLI match the magnification pattern of the lens model, allowing the NDOF to be fixed without suffering the loss in performance of other methods.

\item[(ii)] The use of a fixed source plane pixelization, for example, but not limited to, a square grid of pixels, generally gives rise to a biased set of model parameters with the SPLE lens. We demonstrated this by phase shifting a square grid by fractional pixel widths and showing that each phase gives a significantly different and generally biased lens model. By deriving the pixelization of every set of lens model parameters in an always different and unique way, adaptive SLI naturally explores all systematics associated with data discretization and thus removes all associated biases. We investigated the use of adaptive SLI with a fixed, not random, initialization (see appendix \ref{AppA}), in line with some existing inversion methods and found that the biases not only persisted but were in fact amplified compared to square SLI, although we note that the severity of the effect will vary significantly depending upon the exact implementation.

\end{itemize}

Through its removal of discretization biases, adaptive SLI gives accurate and robust sampling of the lens model posterior probability distribution function with just one non-linear search, something we believe is not possible with existing methods. We demonstrated this by fitting the highly degenerate SPLE to our two simulated images. By design, adaptive SLI copes well with highly degenerate profiles like the SPLE and it therefore offers strong potential for the fitting of more sophisticated profiles, for which parameter degeneracies are even more challenging. For example, models which decompose the mass into a baryonic component and a dark matter component possess a strong degeneracy between the two.

Of course, several methods in the literature use a source pixelization that depends on the lens model. However, as discussed in section \ref{MethodMass}, none use the random initialization of adaptive SLI. In this way, they therefore calculate related pixelizations, restricting them, like square SLI, to a specific subset of possible data discretizations which will therefore not fully remove the discretization effects we have shown. 

A future goal of adaptive SLI is its development into a strong lensing analysis pipeline. The number of detected strong lens systems is presently undergoing a period of acceleration which is set to continue in the foreseeable future with surveys such as DES, LSST and Euclid. Accordingly, robust lens modeling techniques which offer more standalone functionality, reduced user setup and higher efficiency are becoming increasingly sought after. As demonstrated in this work, the robustness of adaptive SLI gives the necessary strong foundation upon which a lens modeling pipeline can be built. 

The biases outlined in this paper, most noticeably those related to data discretization, are just one example of the systematics which can dramatically affect the results of lens modeling. There are many more systematics which have not been fully investigated, not limited to issues such as PSF accuracy, the use of over-simplified parametric lens profiles, the impact of image quality and intrinsic source morphology. As strong lensing inversion methods grow in their maturity and begin to be used as standard across many different astronomical disciplines, it is imperative that such effects are thoroughly explored in the short-term future.

\appendix
\section{Data Discretization With Fixed Initialization}\label{AppA}

The use of random initialization in adaptive SLI ensures that data
discretization is different and unique with each lens model parameter set.
Here, we further investigate this important feature. In particular, we
wish to test the consequences of fixing the initialization of h-cluster centers, since this should give rise to an overall smoother
variation of evidence with lens parameters and therefore aid optimization.

Random initialization occurs because the initial centers from which the
clustering algorithm proceeds are calculated randomly from the input
spatial coordinates (the traced image pixels coordinates see section \ref{MethodNoReg} for more details). Initialization is
then easily fixed by ensuring these initial centers are always the same
every time the clustering algorithm starts. The adaptive pixelization
still continues as normal, but this process behaves very
deterministically. Therefore, nearly identical lens models now
give rise to nearly identical pixelizations and the data discretization
for similar lens models are now related and dependent on the lens model,
as with other methods in the literature.

In terms of fixing the initialization, we define our fixed initial centers
as those given by the final pixelization calculated by adaptive SLI for
each of images 1 and 2. We wish to investigate the result of changing the
initialization, therefore we also calculate a set of eight other random
initialization's corresponding to slight perturbations in the input lens
model. In this way, we end up with nine different sets of cluster centers
which are then fixed for each of the nine corresponding adaptive SLI runs.
This set up is analogous to the phasing shifting of nine grids with square
SLI. The inversion setup parameters (e.g. source resolutions, image
subgridding, regularization) are the same as those used in the main paper.

\begin{figure}
\centering
\includegraphics[width=0.5\textwidth]{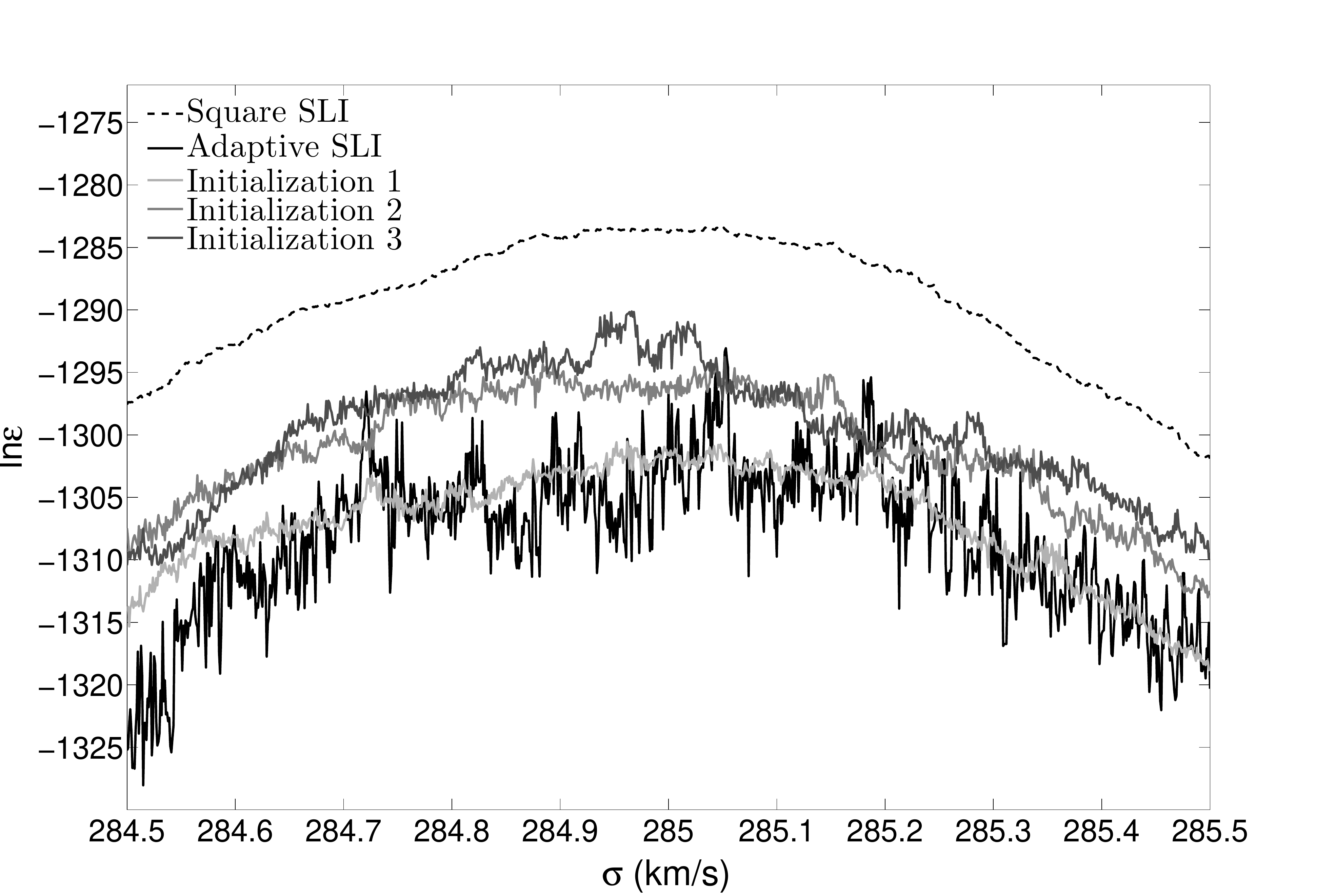}
\caption{Inversion of image 1 using the true lens model except the
velocity dispersion, $\sigma$, which is varied over $\pm 0.5$\,kms$^{-1}$
about the true value. The figure is as figure \ref{figure:LHNoise} except
here we show three adaptive inversions with different but fixed initial
source pixelizations. The figure shows that the resulting variation of
evidence with $\alpha$ is smoother when fixing the source pixelization
compared to allowing it to vary with each lens model iteration (black
line).}
\label{figure:LHNoiseFixed}
\end{figure}

We first ensure that a fixed initialization smooths non-linear parameter
space by repeating the demonstration given in figure \ref{figure:LHNoise},
whereby we fix every parameter to that of the input lens models except
$\sigma$, which we vary over the range 284.5 km/s to 285.5 km/s. The
results are given in figure \ref{figure:LHNoiseFixed} for three of the
nine fixed initializations. As expected, the variation of evidence with
$\sigma$ is much smoother than adaptive SLI when initialization is
randomized. The results of using square and randomized adaptive SLI are
also plotted on figure \ref{figure:LHNoiseFixed} for comparison. This
method acts to smooth parameter space on small scales and will result in
a full non-linear search using significantly fewer iterations to find the
optimal lens model. However, the figure shows that a hint of the return of
discreteness bias since all three evidence curves follow different shapes
and peak at different values of $\sigma$.

\begin{figure*}
\centering
\includegraphics[width=0.97\textwidth]{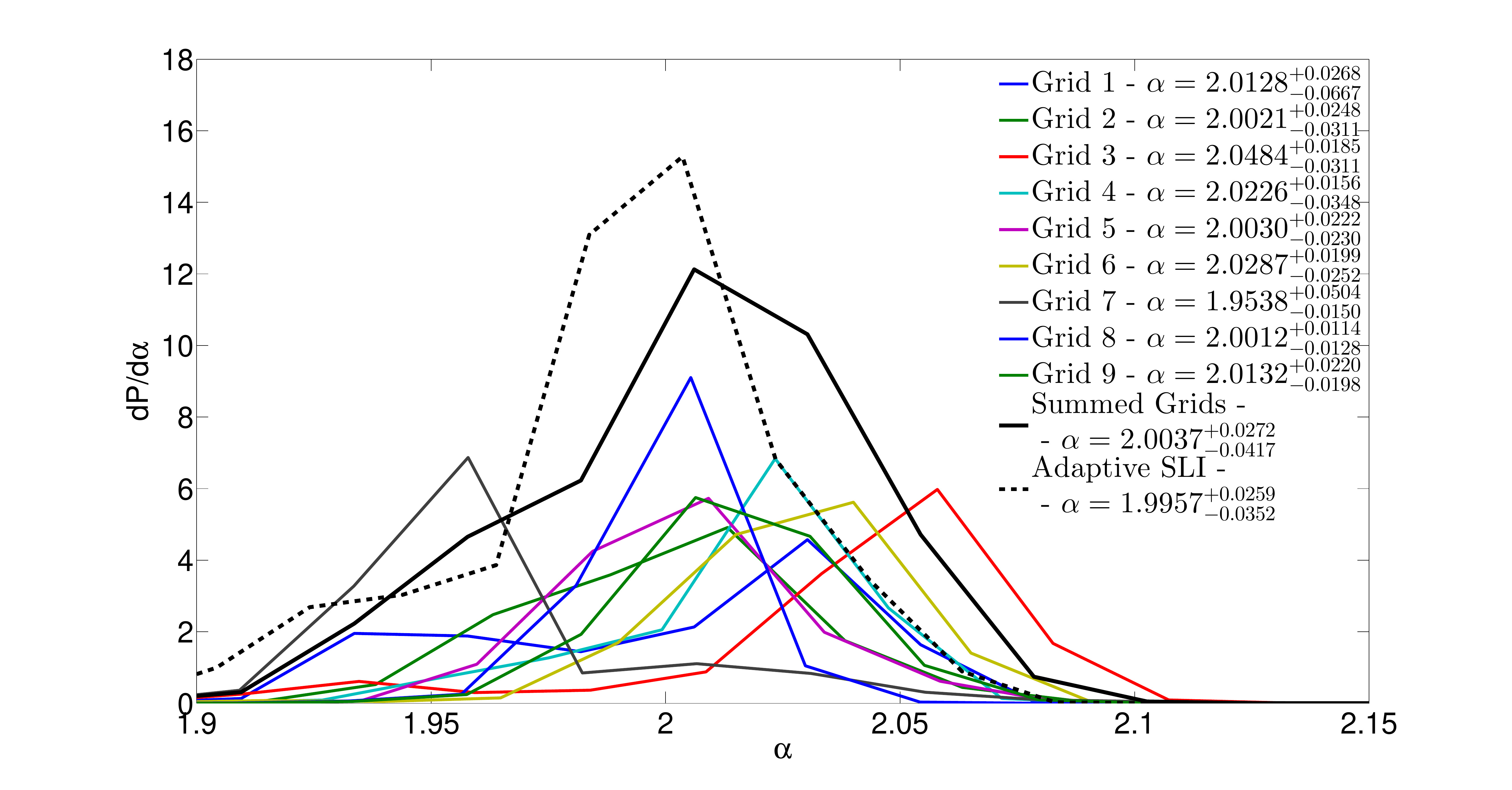}
\includegraphics[width=0.97\textwidth]{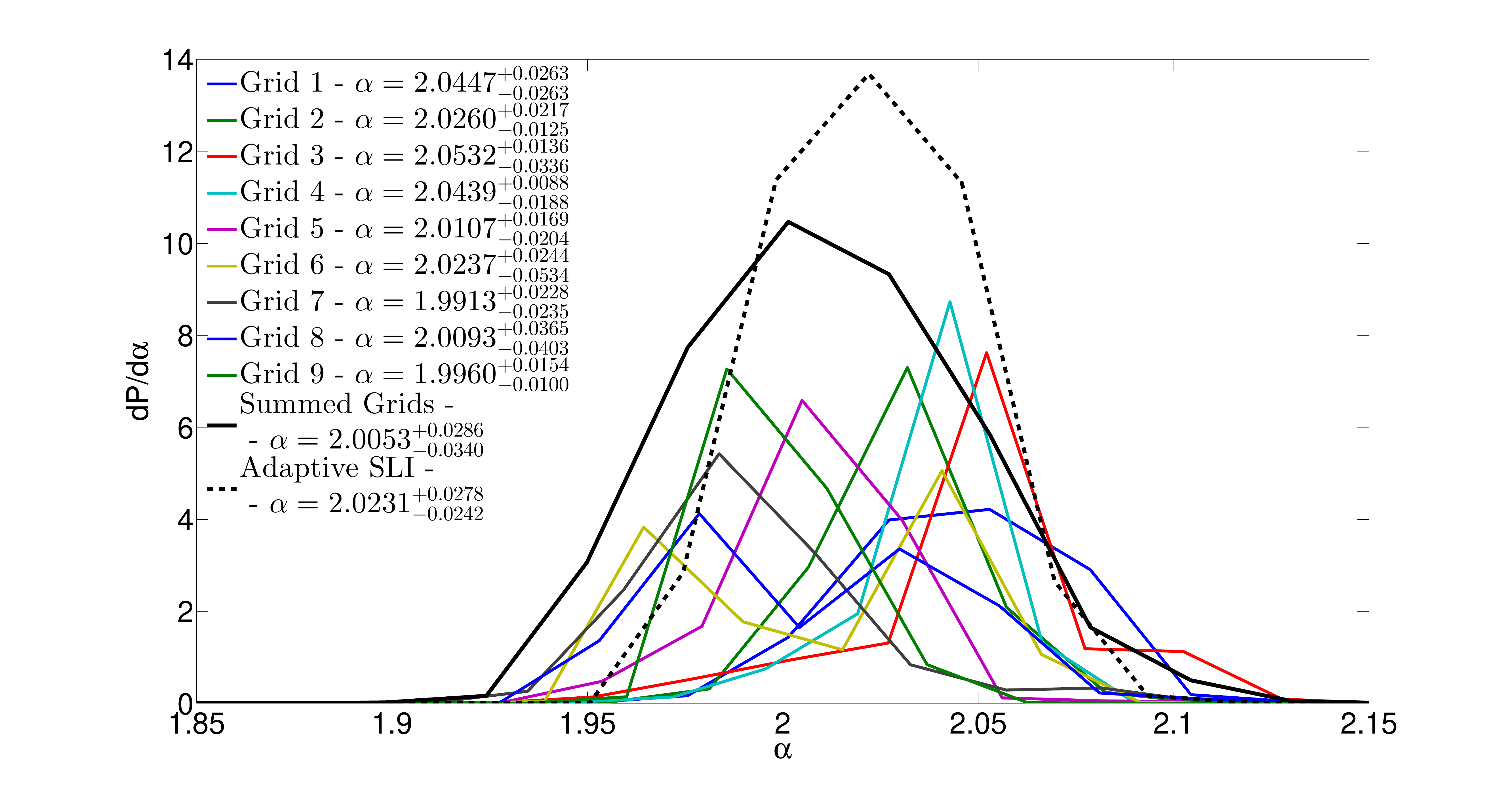}
\caption{Posterior distribution function (PDF) of $\alpha$ obtained by
adaptive SLI with fixed initialization of source pixel cluster centers for
image 1 (top) and image 2 (bottom). The inversion was set up with $4
\times 4$ sub-gridding, an arbitrarily large source plane and 200
clusters. The range of $\alpha$ values show greater inconsistency and
lower accuracy than those in figure \ref{figure:GridAlign_isub4}, showing
that fixed initialization in adaptive SLI reintroduces discreteness biases
and amplifies them compared to square SLI. The thick black line shows the
summed PDF of all fixed grids and the dashed line shows the accurate lens
models calculated using adaptive SLI.}
\label{figure:GridAlign_FixClust}
\end{figure*}

We now perform a full non-linear search for every fixed initialization for
images 1 and 2, as we did when phase shifting with square SLI. The PDFs
are plotted in figure \ref{figure:GridAlign_FixClust} in the same way as
those showing square SLI phase shifts in figure
\ref{figure:GridAlign_isub4}.

Surprisingly, for image 1 discreteness biases are present and infact more severe than those found when phase
shifting square SLI. While square SLI gave mostly accurate and
consistent lens models for image 1, in this case they span a wider range
of $\alpha$ with greater inconsistency, while image 2 remains inaccurate, however with slightly less variation among runs than found with square SLI.
Furthermore, individual PDFs appear more sharply peaked and narrower than
either the summed PDF or that found with adaptive SLI. Therefore the underestimation of errors is present
once again and demonstrates that without full consideration of the data
discretization, biases in errors arise.

Interestingly, for boths images the average $\alpha$ PDF summed over the 9 grids is accurate. 
This is expected, given that this averages over discreteness biases in a similiar fashion to how we present adaptive SLI in the main body of this paper, 
demonstrating that the removal of discretization biases can only be achieved by averaging over multiple source plane discretizations. 
While the method shown in this appendix benefits from a reduction in the number
of iterations required to find each best-fit lens model, we feel this is negated
by the requirement to average over multiple differently initialized grids. We therefore advocate the use of adaptive SLI as presented in the main body of this paper.

These results make the incorrect summed phase shift value of $\alpha = 2.0437$ found using square SLI on image 2 worthy of further consideration.
The nine fixed adaptive SLI grids used in this appendix have no prescribed geometric form governing their source plane pixelization and thus averaging over them explores the possible different forms of source plane discretization. 
On the other hand phase shifted square grids still adhere to the same overall symmetric geometry. 
This biased value of $\alpha$ may then be the result of square SLI not having enough freedom to fully sample different discretizations, leading to a systematic bias in parameter estimation.  
Alternatively it may be a result of the the $N_{pix}$ bias described in \ref{NpixBias}, however this would be surprising given any effect of this bias is expected to be minimal when equation \ref{eqn:evidence} is optimized for every lens model, as was done.

We expect that data discretization biases are present in other inversion
methods within the literature, such as the adaptive square grids of
\cite{Dye2005,Tagore2014}, rectangular grids of
\cite{Collett2014,Suyu2013} and adaptive Voronoi grid of
\cite{Vegetti2009}; none use random initialization like adaptive SLI and
thus source pixelizations are not unique nor unrelated between lens model
iterations. The offset value of $\alpha$ found phasing shifting square SLI on image 2 also suggests these methods may be subject to systematically biased and offset results, even when averaged over multiple discretizations. However the fact this is not repeated in image 1 serves to show that the severity of any such effect depends on many factors
and is therefore hard to predict in specific cases.

We conclude that the random initialization of adaptive SLI is a vital
component to the method, ensuring it can fully explore all systematics
associated with data discretization to ultimately give an unbiased and
accurate lens model using just one non-linear search.

\section{Further Testing of Adaptive Semi-linear Inversion}\label{AppB}

The fixed setup we used in the main body of this paper of 200 source pixels and $4 \times 4$ image subgridding was chosen as a compromise between ensuring sufficient resolution for accurate results and keeping computational run-time and overheads minimal. This does however raise the question of the sensitivity of our results to the setup used. In this appendix, we therefore repeat the fitting of a SPLE to images 1 and 2 for a variety of different setups.

We first repeat the modeling of section \ref{NpixBias} but with a different number of adaptive pixels. In this section, $\lambda$ was optimized for every lens model by maximizing the evidence in equation \ref{eqn:evidence}. The results, which were obtained for 200 adaptive pixels, are repeated in table \ref{table:SrcResEvi} along with additional results for 300 and 400 adaptive pixels for comparison.

The effect of changing source plane resolution is minimal; all models are fully consistent with one another. This is perhaps not surprising, since the increase in source plane resolution is compensated for by an increase in regularization which correlates source pixels thus decreasing the overall effective source plane resolution. These results show that adaptive SLI retains accuracy while modeling a degenerate lens profile like the SPLE, even with a relatively high ratio of image pixels to source pixels (200 source pixels corresponds to a ratio 11.6 for image 1, 6.74 for image 2). 

The fourth and fifth columns of \ref{table:SrcResEvi} show the result of increasing the image subgridding to $8 \times 8$. Again, all results remain fully consistent with one another. There appears here to be little benefit in increasing the image subgridding beyond a threshold value, although its importance will be more significant for sources which show more irregularity and structure than images 1 and 2.

\begin{table*}
\begin{tabular}{ l | l | l | l | l } 
\multicolumn{1}{p{1.8cm}|}{\centering Source Plane Resolution} 
& \multicolumn{1}{p{2.5cm}|}{\centering Image 1 \\ Sub-gridding $4 \times 4$} 
& \multicolumn{1}{p{2.5cm}|}{\centering Image 2 \\ Sub-gridding $4 \times 4$} 
& \multicolumn{1}{p{2.5cm}|}{\centering Image 1 \\ Sub-gridding $8 \times 8$}
& \multicolumn{1}{p{2.5cm}}{\centering Image 2 \\ Sub-gridding $8 \times 8$}
\\ \hline
& & & & \\[-4pt]
 200 & $1.9958 ^{+0.0258} _{-0.0354}$ & $2.0231 ^{+0.0278} _{-0.0242}$ & $1.9898 ^{+0.0313} _{-0.0335}$ & $2.0247 ^{+0.0191} _{-0.0428}$ \\ [2pt]
 300 & $2.0135 ^{+0.0196} _{-0.0317}$ & $2.0157 ^{+0.0121} _{-0.0180}$ & $1.9970 ^{+0.0260} _{-0.0294}$ & $2.0362 ^{+0.0301} _{-0.0146}$ \\ [2pt]
 400 & $1.9872 ^{+0.0195} _{-0.0328}$ & $2.0333 ^{+0.0167} _{-0.0224}$ & $1.9918 ^{+0.0248} _{-0.0304}$ & $2.0297 ^{+0.0225} _{-0.0443}$ \\ [4pt]
\end{tabular}
\caption{The values of $\alpha$ estimated using adaptive SLI with varying source resolutions and image subgridding. All runs set the regularization coefficient for each lens model by maximizing equation \ref{eqn:evidence} to set $\lambda$. The second and third columns are results for images 1 and 2 with image sub-gridding $4 \times 4$ and the fourth and fifth increase sub-gridding to $8 \times 8$, with all columns showing runs with 200, 300 and 400 adaptive source plane pixels.}
\label{table:SrcResEvi}
\end{table*}

We next repeat the modeling of section \ref{FixReg}, where we used the merit function $G = \chi^2 + \lambda G_L$ with $\lambda$ fixed at an optimal value calculated once prior to inversion to improve efficiency. As such, here, we increase the source plane resolution to give an image:source plane pixel number ratio of 6 (388 source pixels for image 1, 225 image 2) and 3 (776 source pixels for image 1, 451 image 2). All runs use image subgridding of $4 \times 4$. 

The results are given in table \ref{table:SrcResNoEvi}. Once again, all results remain fully consistent with one another, showing that our choice of source plane resolution also has minimal impact on modeling results. This further strengthens our argument that this faster, less expensive modeling technique is accurate.

\begin{table*}
\begin{tabular}{ l | l | l  | l | l } 
 \multicolumn{1}{p{2.5cm}|}{\centering Source pixel / Image pixel ratio} 
& \multicolumn{1}{p{1.8cm}|}{\centering Source Plane Resolution (Image 1)} 
& \multicolumn{1}{p{1.8cm}|}{\centering Source Plane Resolution (Image 2)} 
& \multicolumn{1}{p{2.5cm}|}{\centering Image 1 \\ Sub-gridding $4 \times 4$} 
& \multicolumn{1}{p{2.5cm}}{\centering Image 2 \\ Sub-gridding $4 \times 4$} 
\\ \hline
& & & & \\[-4pt]
N/A & 200 & 200 & $1.9909 ^{+0.0366} _{-0.0321}$ & $2.0225 ^{+0.0297} _{-0.0269}$ \\ [2pt]
6   & 388 & 225 & $1.9774 ^{+0.0386} _{-0.0117}$ & $2.0234 ^{+0.0225} _{-0.0318}$ \\ [2pt]
3   & 776 & 541 & $1.9851 ^{+0.0301} _{-0.0221}$ & $2.0375 ^{+0.0388} _{-0.0442}$ \\ [4pt]
\end{tabular}
\caption{The values of $\alpha$ estimated using adaptive SLI with varying source resolutions. All runs set the regularization coefficient by maximizing equation \ref{eqn:evidence} pre-inversion once to set $\lambda$, then minimizing the merit function $G = \chi^2 + \lambda G_L$. The fourth and fifth columns are results for images 1 and 2 with image sub-gridding $4 \times 4$.}
\label{table:SrcResNoEvi}
\end{table*}

Finally we explore the effect of changing the intrinsic source light distribution. The use of a Gaussian source gives a relatively flat intrinsic light distribution. It is important to test our regularization scheme on one which is both steeper and more centrally peaked. This is an issue faced when modeling the extended host-galaxy around a strongly lensed, point source quasar, for which current analysis methods simply mask out the brightest regions of the image to ensure the regularization is optimized sensibly \cite{Suyu2012}. To test this, we use the source configuration used to create images 1 and 2 but with an $n=3.5$ Sersic profile surrounded by an extended exponential light profile, representing a bulge dominated spiral galaxy (S. Bamford, private communication).  

\begin{table*}
\begin{tabular}{ l | l | l | l | l | l } 
 \multicolumn{1}{p{2.5cm}|}{\centering Source pixel / Image pixel ratio} 
& \multicolumn{1}{p{1.8cm}|}{\centering Source Plane Resolution (Image 1)} 
& \multicolumn{1}{p{1.8cm}|}{\centering Source Plane Resolution (Image 2)} 
& \multicolumn{1}{p{2.5cm}|}{\centering Image 1 \\ Sub-gridding $4 \times 4$} 
& \multicolumn{1}{p{2.5cm}|}{\centering Image 2 \\ Sub-gridding $4 \times 4$} 
& \multicolumn{1}{p{2.5cm}}{\centering Image 2 \\ Sub-gridding $8 \times 8$} 
\\ \hline
& & & & \\[-4pt]
N/A & 400 & 400 & $2.0043 ^{+0.0332} _{-0.0411}$ & $2.0794 ^{+0.0315} _{-0.0460}$ & N/A \\ [2pt]
8   & 474 & 546 & $1.9877 ^{+0.0315} _{-0.0426}$ & $2.0821 ^{+0.0429} _{-0.0493}$ & N/A \\ [2pt]
6   & 633 & 728 & $1.9878 ^{+0.0309} _{-0.0219}$ & $2.0615 ^{+0.0345} _{-0.0772}$ & $2.0433 ^{+0.0294} _{-0.0419}$\\ [2pt]
4   & N/A & 1274& N/A                            & $2.0501 ^{+0.0309} _{-0.0665}$ & $2.0397 ^{+0.0340} _{-0.0447}$\\ [4pt]
\end{tabular}
\caption{The values of $\alpha$ estimated using adaptive SLI for remakes of images 1 and 2 which use a Sersic Bulge + Exponential light distribution. The first row the regularization coefficient is set by maximizing equation \ref{eqn:evidence} for every lens model, whereas allow rows below do this once pre-inversion with the merit function $G = \chi^2 + \lambda G_L$.}
\label{table:SrcBD}
\end{table*}

The results of this analysis are given in table \ref{table:SrcBD}. The exponential light representing the disk leads to a thicker Einstein ring. We therefore extended the image masks for both images 1 and 2, resulting in an increase in the number of image pixels used and thus resulting in a higher source resolution for the same image:source pixel number ratios. For the re-analysis of image 1, changing the source from a Gaussian has little effect. However, the same is not true of image 2 which over-estimates $\alpha$ for low source plane resolution setups.

For image 2, the source's Sersic bulge, where the light profile is steepest, is located above the top right cusp (see figure \ref{figure:SynIm}). Thus, the steepest region of the light profile is only doubly imaged. A high source plane resolution is therefore required to accurately reconstruct this steep light profile and if this is insufficient regularization will smooth over it, giving an inaccurate lens model. 

The high resolution modeling runs in table \ref{table:SrcBD} still show a tendency to over estimate $\alpha$ for image 2, although the modeling errors increase to account for this. As shown in the final column of \ref{table:SrcBD}, increasing subgridding further reduces the values of $\alpha$ estimated. The majority of spiral galaxies observed, even those with a bulge, show much shallower light profiles than our test simulation here \citep{Bruce2012} and thus this is likely not a concern for the majority of observed strong lenses. However it is still an issue worthy of future investigation, especially if one is primarily interested in small-scale features in the reconstructed source (e.g. bulges, bars, star forming knots).

\section*{Acknowledgements}

JN acknowledges support from STFC and the University of Nottingham. SD
acknowledges support from the Midlands Physics Alliance. We are grateful for access to the University of Nottingham High Performance Computing Facility.

\bibliography{Master}

\begin{thebibliography}{}

\bibitem[\protect\citeauthoryear{Auger, Treu, Bolton \& Koopmans}{Auger
  et~al.}{2009}]{Auger2009}
Auger M.,  Treu T.,  Bolton A.,    Koopmans R. G.~L.,  2009, ApJ, 705, 1099

\bibitem[\protect\citeauthoryear{Barnabe, Czosoke, Koopmans \& Treu}{Barnabe
  et~al.}{2009}]{Barnebe2009}
Barnabe M.,  Czosoke O.,  Koopmans L.,    Treu T.,  2009, MNRAS, 399, 21

\bibitem[\protect\citeauthoryear{Bolton, Brownstein, Kochanek \& Shu}{Bolton
  et~al.}{2012}]{Bolton2012}
Bolton A.,  Brownstein J.,  Kochanek C.,    Shu Y.,  2012, ApJ, 757, 82

\bibitem[\protect\citeauthoryear{Bolton, Burles, Koopmans, Treu \&
  Moustakas}{Bolton et~al.}{2006}]{Bolton2006}
Bolton A.,  Burles S.,  Koopmans L.,  Treu T.,    Moustakas L.,  2006, ApJ,
  628, 703

\bibitem[\protect\citeauthoryear{Bolton, Burles, Koopmans, Treu \&
  Moustakas}{Bolton et~al.}{2008}]{Bolton2008}
Bolton A.,  Burles S.,  Koopmans L.,  Treu T.,    Moustakas L.,  2008, ApJ,
  682, 964

\bibitem[\protect\citeauthoryear{Bolton, Treu, Koopmans, Gavazzi \&
  Moustakas}{Bolton et~al.}{2008}]{Bolton2008b}
Bolton A.,  Treu T.,  Koopmans L.,  Gavazzi R.,    Moustakas L.,  2008, ApJ,
  684, 248

\bibitem[\protect\citeauthoryear{Brownstein, Bolton, Schlegel \&
  Eisenstein}{Brownstein et~al.}{2012}]{Brownstein2012}
Brownstein J.,  Bolton A.,  Schlegel D.,    Eisenstein D.,  2012, ApJ, 744, 41

\bibitem[\protect\citeauthoryear{Bruce, Dunlop, Cirasuolo \& Mclure}{Bruce
  et~al.}{2012}]{Bruce2012}
Bruce V.,  Dunlop J.,  Cirasuolo M.,    Mclure R.,  2012, MNRAS, 427, 1666

\bibitem[\protect\citeauthoryear{Collett \& Auger}{Collett \&
  Auger}{2014}]{Collett2014}
Collett T.,  Auger M.,  2014, MNRAS, 443, 969

\bibitem[\protect\citeauthoryear{Dutton \& Treu}{Dutton \&
  Treu}{2014}]{Dutton2014}
Dutton A.,  Treu T.,  2014, MNRAS, 438, 3594

\bibitem[\protect\citeauthoryear{Dye, Evans, Belokurov, Warren \& Hewett}{Dye
  et~al.}{2008}]{Dye2008}
Dye S.,  Evans N.,  Belokurov V.,  Warren S.,    Hewett P.,  2008, MNRAS, 000,
  1

\bibitem[\protect\citeauthoryear{Dye, Negrello, Hopwood \& Nightingale}{Dye
  et~al.}{2014}]{Dye2014}
Dye S.,  Negrello M.,  Hopwood R.,    Nightingale J.,  2014, MNRAS, 000, 1

\bibitem[\protect\citeauthoryear{Dye \& Warren}{Dye \& Warren}{2005}]{Dye2005}
Dye S.,  Warren S.,  2005, ApJ, 623, 31

\bibitem[\protect\citeauthoryear{Feroz \& Hobson}{Feroz \&
  Hobson}{2007}]{Feroz2007}
Feroz F.,  Hobson M.,  2007, MNRAS, 384, 449

\bibitem[\protect\citeauthoryear{Feroz, Hobson \& Bridges}{Feroz
  et~al.}{2009}]{Feroz2009}
Feroz F.,  Hobson M.,    Bridges M.,  2009, MNRAS, 398, 1601

\bibitem[\protect\citeauthoryear{Gavazzi, Treu, Marshall \& Brault}{Gavazzi
  et~al.}{2012}]{Gavazzi2012}
Gavazzi R.,  Treu T.,  Marshall P.,    Brault F.,  2012, ApJ, 761-2, 170

\bibitem[\protect\citeauthoryear{Hartigan \& Wong}{Hartigan \&
  Wong}{1979}]{Hartigan1979}
Hartigan J.,  Wong M.,  1979, JRSS, 28.1, 100

\bibitem[\protect\citeauthoryear{Keeton}{Keeton}{2002}]{Keeton2002}
Keeton C.,  2002, ARXIV, preprint

\bibitem[\protect\citeauthoryear{Koopmans, Bolton, Treu \& Czoske}{Koopmans
  et~al.}{2009}]{Koopmans2009}
Koopmans L.,  Bolton A.,  Treu T.,    Czoske O.,  2009, ApJ, 703, 51

\bibitem[\protect\citeauthoryear{Koopmans, Treu, Bolton, Burles \&
  Moustakas}{Koopmans et~al.}{2006}]{Koopmans2006}
Koopmans L.,  Treu T.,  Bolton A.,  Burles S.,    Moustakas L.,  2006, ApJ,
  649-2, 599

\bibitem[\protect\citeauthoryear{Lagattuta \& Vegetti}{Lagattuta \&
  Vegetti}{2012}]{Lagattuta2012}
Lagattuta D.,  Vegetti S.,  2012, MNRAS, 424, 2800

\bibitem[\protect\citeauthoryear{Oguri \& Marshall}{Oguri \&
  Marshall}{2010}]{Oguri2010}
Oguri M.,  Marshall P.,  2010, MNRAS, 405, 2579

\bibitem[\protect\citeauthoryear{Oguri, Rusu \& Falco}{Oguri
  et~al.}{2014}]{Oguri2014}
Oguri M.,  Rusu C.,    Falco E.,  2014, MNRAS, 439, 2494

\bibitem[\protect\citeauthoryear{Schneider \& Sluse}{Schneider \&
  Sluse}{2013}]{Schneider2013}
Schneider P.,  Sluse D.,  2013, A\&A, 559, A37

\bibitem[\protect\citeauthoryear{Skilling}{Skilling}{2006}]{Skilling2006}
Skilling J.,  2006, Bayesian Analysis., 1, 833

\bibitem[\protect\citeauthoryear{Sluse \& Schneider}{Sluse \&
  Schneider}{2013}]{Sluse2013}
Sluse D.,  Schneider P.,  2013, A\&A, 564, A103

\bibitem[\protect\citeauthoryear{Sonnenfeld, Gavazzi, Suyu, Treu \&
  Marshall}{Sonnenfeld et~al.}{2013}]{Sonnenfeld2013b}
Sonnenfeld A.,  Gavazzi R.,  Suyu S.,  Treu T.,    Marshall P.,  2013, ApJ,
  777, 97

\bibitem[\protect\citeauthoryear{Sonnenfeld, Treu, Gavazzi, Suyu \&
  Marshall}{Sonnenfeld et~al.}{2013}]{Sonnenfeld2013c}
Sonnenfeld A.,  Treu T.,  Gavazzi R.,  Suyu S.,    Marshall P.,  2013, ApJ,
  777, 98

\bibitem[\protect\citeauthoryear{Suyu}{Suyu}{2012}]{Suyu2012}
Suyu S.,  2012, MNRAS, 426, 868

\bibitem[\protect\citeauthoryear{Suyu, Auger, Hilbert \& Marshall}{Suyu
  et~al.}{2013}]{Suyu2013}
Suyu S.,  Auger M.,  Hilbert S.,    Marshall P.,  2013, ApJ, 766 - 2, 70

\bibitem[\protect\citeauthoryear{Suyu, Marshall, Hobson \& Blandford}{Suyu
  et~al.}{2006}]{Suyu2006}
Suyu S.,  Marshall P.,  Hobson M.,    Blandford R.,  2006, MNRAS, 371, 983

\bibitem[\protect\citeauthoryear{Tagore \& Keeton}{Tagore \&
  Keeton}{2014}]{Tagore2014}
Tagore A.~S.,  Keeton C.~R.,  2014, MNRAS, 371, 983

\bibitem[\protect\citeauthoryear{Treu, Dutton, Auger \& Marshall}{Treu
  et~al.}{2011}]{Treu2011}
Treu T.,  Dutton A.,  Auger M.,    Marshall P.,  2011, MNRAS, 417, 1601

\bibitem[\protect\citeauthoryear{Treu \& Koopmans}{Treu \&
  Koopmans}{2002}]{Treu2002}
Treu T.,  Koopmans L.,  2002, ApJ, 575, 87

\bibitem[\protect\citeauthoryear{Treu \& Koopmans}{Treu \&
  Koopmans}{2004}]{Treu2004}
Treu T.,  Koopmans L.,  2004, ApJ, 611, 739

\bibitem[\protect\citeauthoryear{Vegetti \& Koopmans}{Vegetti \&
  Koopmans}{2009}]{Vegetti2009}
Vegetti S.,  Koopmans L.,  2009, MNRAS, 392, 945

\bibitem[\protect\citeauthoryear{Warren \& Dye}{Warren \&
  Dye}{2003}]{Warren2003}
Warren S.,  Dye S.,  2003, ApJ, 590, 673

\end{thebibliography}

\label{lastpage}

\end{document}